\documentclass[12pt,preprint,referee]{aastex}

   \usepackage{color}

\usepackage{gensymb}

\newcommand{\be} {\begin{equation}}

\def\psrj {PSR\,J0007$+$7303\/}
\def\qso {S5\,0016+73\/}
\usepackage{gensymb}

\newcommand{\fermi}{{\em Fermi}}

\newcommand{\bc}{\begin{center}}
\newcommand{\ec}{\end{center}}
\def\ltsima{$\; \buildrel < \over \sim \;$}
\def\lsim{\lower.5ex\hbox{\ltsima}}
\def\loe{\lower.5ex\hbox{\ltsima}}
\def\gtsima{$\; \buildrel > \over \sim \;$}
\def\gsim{\lower.5ex\hbox{\gtsima}}
\def\goe{\lower.5ex\hbox{\gtsima}}

\def\ltsima{$\; \buildrel < \over \sim \;$}
\def\lsim{\lower.5ex\hbox{\ltsima}}
\def\loe{\lower.5ex\hbox{\ltsima}}
\def\gtsima{$\; \buildrel > \over \sim \;$}
\def\gsim{\lower.5ex\hbox{\gtsima}}
\def\goe{\lower.5ex\hbox{\gtsima}}

\def\ergscm2 {erg\,s$^{-1}$cm$^{-2}$}

\def\cm2 {cm$^{-2}$}

\begin{document}
\title{Gamma-ray emission from \psrj\ using 7 years of {\em Fermi} Large Area Telescope observations }

\author{Jian Li\altaffilmark{1}, Diego F. Torres\altaffilmark{1,2}, Emma de O\~na Wilhelmi\altaffilmark{1}, \\ Nanda Rea\altaffilmark{1,3}, Jonatan Martin\altaffilmark{1}}
\altaffiltext{1}{Institute of Space Sciences (CSIC--IEEC), Campus UAB, Carrer de Magrans s/n, 08193 Barcelona, Spain}
\altaffiltext{2}{Instituci\'o Catalana de Recerca i Estudis Avan\c{c}ats (ICREA), E-08010 Barcelona, Spain}
\altaffiltext{3}{Astronomical Institute ``Anton Pannekoek", University of Amsterdam, Postbus 94249, NL-1090-GE Amsterdam, The Netherlands}

\begin{abstract}

Based on more than seven years of \emph{Fermi} {Large Area Telescope (LAT)} Pass 8 data, we report on a detailed analysis of the bright gamma-ray pulsar (PSR) J0007$+$7303\/.
We confirm that \psrj\ is significantly detected as a point source also during the off-peak phases {with a TS value of 262  ($\sim$ 16 $\sigma$)}.
%
%
{In the description of \psrj\/ off-peak spectrum, a power law with an exponential cutoff at 2.7$\pm$1.2$\pm$1.3 GeV (the first/second uncertainties correspond to statistical/systematic errors) is preferred over a single power law at a level of 3.5 $\sigma$. The possible existence of a cutoff hints at a magnetospheric origin of the emission.}
In addition, no extended gamma-ray emission is detected compatible with either the supernova remnant (CTA~1) or the very high energy ($>$ 100 GeV) pulsar wind nebula.
A flux upper limit of 6.5$\times$10$^{-12}$ erg cm$^{-2}$ s$^{-1}$ in the 10-300 GeV energy range is reported, for an extended source assuming the morphology of the VERITAS detection.
During on-peak phases, a sub-exponential cutoff is significantly preferred ($\sim$ 11 $\sigma$) for representing the spectral energy distribution, both in the phase-averaged and in the phase-resolved spectra.
Three glitches are detected during the observation period and we found {no flux variability at the time of the glitches} or in the long-term behavior.
We also report the discovery of a previously unknown gamma-ray source in the vicinity of PSR J0007+7303, Fermi J0020+7328, which we associate with the $z=1.781$ quasar \qso\/.
A concurrent analysis of this source is needed to correctly characterize the behavior of CTA~1 and it is also presented in the paper.

\end{abstract}


\section{INTRODUCTION}
\label{intro}

\psrj\/ is a $\sim$ 316 ms gamma-ray pulsar discovered by the \emph{Fermi} Large Area Telescope (LAT) in a blind search (Abdo et al. 2008).
Using the timing ephemeris from the LAT, X-ray pulsations from \psrj\/ were detected by \emph{XMM-Newton} (Lin et al. 2010; Caraveo et al. 2010).
Deep searches for optical and radio counterparts of \psrj\/ revealed none (Halpern et al. 2004; Mignani et al. 2013), leading to the characterization of \psrj\/ as a {radio-quiet} gamma-ray pulsar similar to Geminga (Bertsch et al. 1992) and PSR J1836+5925 (Halpern, Camilo \& Gotthelf 2007; Abdo et al. 2010; Lin et al. 2014).

 \psrj\ is one of the brightest pulsars in The Second \emph{Fermi} Large Area Telescope Catalog of Gamma-Ray Pulsars (Abdo et al. 2013, 2PC hereafter), providing enough statistics to investigate spectral and timing features and flux variability in detail.
\psrj\/ is associated with the composite supernova remnant CTA~1 (SNR; G119.5+10.2),  discovered by Harris \& Roberts (1960).
CTA~1 possesses a large radio shell that is incomplete towards the north-west (Pineault et al. 1993).
\emph{ASCA} and \emph{ROSAT} observations revealed a central filled SNR with emission extending to the radio shell (Seward et al. 1995).
\emph{Chandra} observations resulted in the detection of a pulsar wind nebula (PWN) and a jet-like structure (Halpern et al. 2004).
The age of CTA~1 is estimated to be around 10 kyr (Pineault et al 1993; Slane et al. 1997; 2004) and the distance is estimated to be 1.4 $\pm$ 0.3 kpc based on the associated H\,{\sc i} shell (Pineault 1997).

The CTA~1 complex was established as an extended gamma-ray source above 500 GeV by VERITAS (Aliu et al. 2013).
The extended morphology detected by VERITAS was approximated by a two-dimensional Gaussian of semi-major (semi-minor) axis of 0$\fdg$30$\pm$0$\fdg$03 (0$\fdg$24$\pm$0$\fdg$03).
The TeV photon origin was proposed to be the PWN associated with \psrj\ (Aliu et al. 2013).
With two years of \emph{Fermi}-LAT observations, the off-peak emission of \psrj\ appeared to be extended and the morphology was fitted with a disk of radius 0$\fdg$7$\pm$0$\fdg$3 at 95\% confidence level.
Given the extension and spectral shape derived with the two-year statistics, the emission was proposed to be associated with CTA~1 (Abdo et al. 2012).

In this paper, we report further analysis of \psrj\ and its related SNR CTA~1 using more than seven years of \emph{Fermi}-LAT data and the newest response functions.

\section{OBSERVATIONS}
\label{obs}

The \emph{Fermi}-LAT data used for this paper covers 88 months, from 2008 August 4 (MJD 54682) to 2015 December 14 (MJD 57370), greatly extending the two years of data coverage reported in Abdo et al. (2012) and the three years of coverage of the 2PC.
The LAT is described in Atwood et al. (2009).
The analysis of \emph{Fermi}-LAT data was performed using the \emph{Fermi} Science Tools\footnote{\url{http://fermi.gsfc.nasa.gov/ssc/}},
10-00-05 release.
Events from the ``Pass 8'' event class were selected.
The ``P8R2 V6 Clean'' instrument response functions (IRFs) were used in the analysis.
All gamma-ray photons within an energy range of 0.1--300 GeV and a circular region of interest (ROI) of 10$\degree$ radius centered on \psrj\/ were considered.
To reject contaminating gamma rays from the Earth's limb, we selected events with a zenith angle $<$ 90$\degree$.

The spectral results presented in this work were calculated by performing a binned maximum likelihood fit (30 bins in the 0.1--300 GeV range) using the Science Tool \emph{gtlike}.
The spectral-spatial model constructed to perform the likelihood analysis includes Galactic and isotropic diffuse emission components (``gll\_iem\_v06.fits", Acero et al. 2016, and ``iso\_P8R2\_CLEAN\_V6\_v06.txt", respectively\footnote{\url{http://fermi.gsfc.nasa.gov/ssc/data/access/lat/BackgroundModels.html}}) as well as known gamma-ray sources within 15$\degree$ of  \psrj\/, included in the \emph{Fermi} LAT Third Source Catalog (Acero et al. 2015, 3FGL hereafter).
The spectral parameters and positions were fixed to the catalog values, except for the sources within 3$\degree$ of our target.
For these latter sources, the spectral parameters were left free.
In the phased analysis, the prefactor parameters {were scaled to the {relative} width of the phase interval.}
The test statistic (TS) was employed to evaluate the significance of the gamma-ray fluxes coming from the sources.
The Test Statistic is defined as TS=$-2 \ln (L_{max, 0}/L_{max, 1}) $, where $L_{max, 0}$ is the maximum likelihood value for a model without an additional source (the ``null hypothesis") and $L_{max, 1}$ is the maximum likelihood value for a model with the additional source at a specified location.
The larger the value of TS, the less likely the null hypothesis is correct (i.e., a significant gamma-ray excess lies on the tested position) and the square root of the TS is approximately equal to the detection significance of a given source.

To search for the possible extension of \psrj\ in the off-peak gamma-ray emission, we followed the method of Lande et al. (2012).
The source is assumed to be spatially extended with a symmetric disk model and we fitted its position and extension with the \textit{Pointlike} analysis package (Kerr 2010). 
The extension significance was defined as TS$_{ext}$=2($\ln L_{disk}$-$\ln L_{point}$), in which $L_{disk}$ and $L_{point}$ were the \textit{gtlike} global  likelihood of the point source and the extended source hypotheses, respectively.
%
The TS maps (i.e., maps of the value of TS for trial positions of an additional source) in this paper are produced with \textit{Pointlike}.

The systematic errors have been estimated similarly to other {\it Fermi}-LAT reports, by repeating the analysis using modified IRFs that bracket the effective area (Ackermann et al. 2012), and artificially changing the normalization of the Galactic diffuse model  by $\pm$6\% (2PC).
The energy dispersion is not considered in the data analysis, which may be important below 100 MeV but not expected to produce significant changes in the 100 MeV--300 GeV energy range considered here.

For the \emph{Swift}/XRT data included in our analysis, we selected Photon Counting (PC) mode\footnote{\url{https://www.swift.psu.edu/xrt/software.html}} data with event grades 0--12 (Burrows et al. 2005).
Source events were accumulated within a circular region with a radius of 30 pixels (1 pixel = 2.36 arcsec).
Background events were accumulated within a circular, source-free region with a radius of 60 pixels.
Exposure maps were generated with the task XRTEXPOMAP.
Ancillary response files were generated with XRTMKARF, which accounts for different extraction regions, vignetting, and point-spread function (PSF) corrections.
We analyzed the \emph{Swift}/XRT 0.3--10 keV data using HEAsoft version 6.14\footnote{\url{http://heasarc.nasa.gov/lheasoft/}}, whereas the spectral fitting was performed using XSPEC V.12.8.1.

\section{Off-peak and on-peak phase selection}
\label{results}

\begin{table*}{}
\centering
\scriptsize
\caption{Timing ephemeris of \psrj\/.}

\begin{tabular}{ll}
\hline
\hline
                                   \\
Parameter & Value   \\
\\
\hline\hline                 
\\

Pulsar name &                \psrj\              \\

R.A. (J2000, Halpern et al. 2004) &   00:07:01.56   \\

 Decl. (J2000, Halpern et al. 2004) &   $+$73:03:08.1   \\

MJD range &         54686.16 -- 57370.50       \\

Pulse frequency, $\nu$ (s$^{-1}$) &     3.1658208(2)                         \\

First derivative of pulse frequency, $\dot{\nu}$ (s$^{-2}$) &    -3.59(1)$\times$10$^{-12}$        \\

Second derivative of pulse frequency, $\ddot{\nu}$ (s$^{-3}$) &    2.92$\times$10$^{-22}$ $\dag$        \\

Epoch of frequency determination (MJD) &  54952                           \\

Glitch Epoch 1 (MJD) &           54952.92239                   \\
Glitch Epoch 2 (MJD) &           55463.89923                    \\
Glitch Epoch 3 (MJD) &           56369.56142                   \\



\\

\hline\hline                 
\label{ephemeris}

\tablecomments  {$\dag$: model-predicted value assuming a braking index of 3.}

\end{tabular}
\end{table*}

\begin{center}
\begin{figure*}
\centering
\includegraphics[angle=0, scale=0.6]{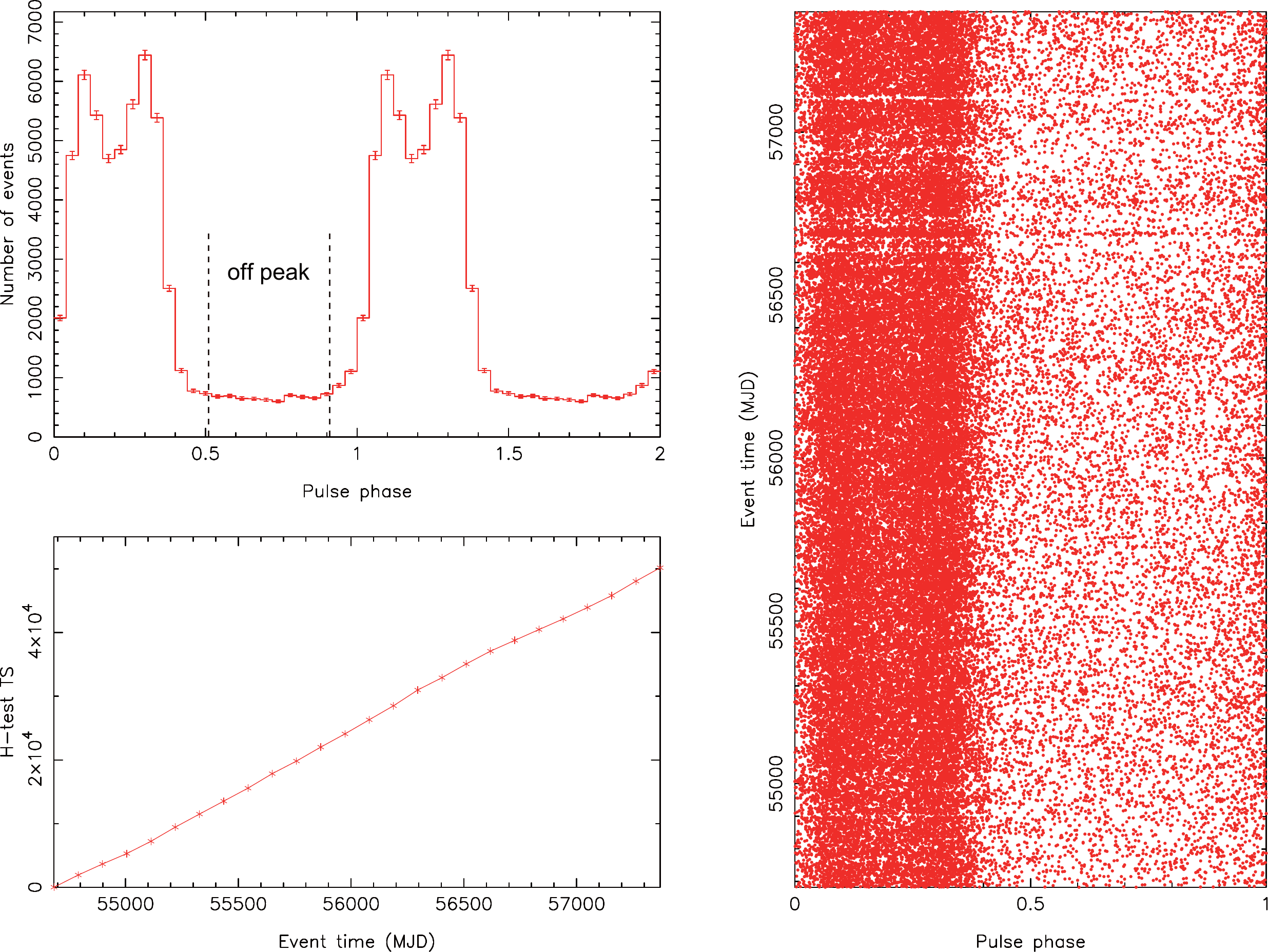}
\caption{\psrj\ timing results from \emph{Tempo2} with the {\it Fermi} plug-in.
Top-left panel: phase histogram of the analyzed gamma-ray data.
%
%
Two full rotations are shown for clarity.
{The off-peak phase interval as determined from a Bayesian block analysis} is $\phi$=0.511$-$0.909 {(the region indicated by the black dashed lines)} while the on-peak phase {interval} is $\phi$=0.0$-$0.511 and  $\phi$=0.909$-$1.0.
Bottom-left panel: H-test significance (TS) as a function of time.
Right panel: pulse phase for each gamma-ray event versus time.}
\label{tempo2}
\end{figure*}
\end{center}

We selected photons from \psrj\ within a radius of 1$\fdg$2 and a minimum energy of 200 MeV, which maximized the H-test statistics { (de Jager et al. 1989; de Jager \& B$\ddot{u}$sching 2010)}.
Adopting the most current ephemeris for \psrj\/ which includes three glitches (M. Kerr 2016, private communication, Table 1), 
we assigned pulsar rotational phases to each gamma-ray photon that passed the selection criteria, using \emph{Tempo2} (Hobbs et al. 2006) with the {\it Fermi} plug-in (Ray et al. 2011).
{The details of the timing analysis and the full timing parameters will be published in the future (M. Kerr et al. 2016, in preparation).\footnote{The timing model will be made available as usual from \url{http://fermi.gsfc.nasa.gov/ssc/data/access/lat/ephems/}.}
The phase reference in our ephemeris is the same as in the 2PC, so that the profiles could be compared directly.}
We adopted the \emph{Chandra} position from Halpern et al. (2004) since the timing noise of \psrj\/ leads to a much lower precision for the measured gamma-ray position (Kerr et al. 2015).
The large timing noise also does not allow a reasonable measurement of the proper motion and braking index of \psrj\/.
The timing results are shown in Figure \ref{tempo2}.
The H-test {TS} increases linearly with time, indicating that the timing ephemeris is valid for the entire data coverage.

\begin{center}
\begin{figure*}
\centering
\includegraphics[scale=0.6]{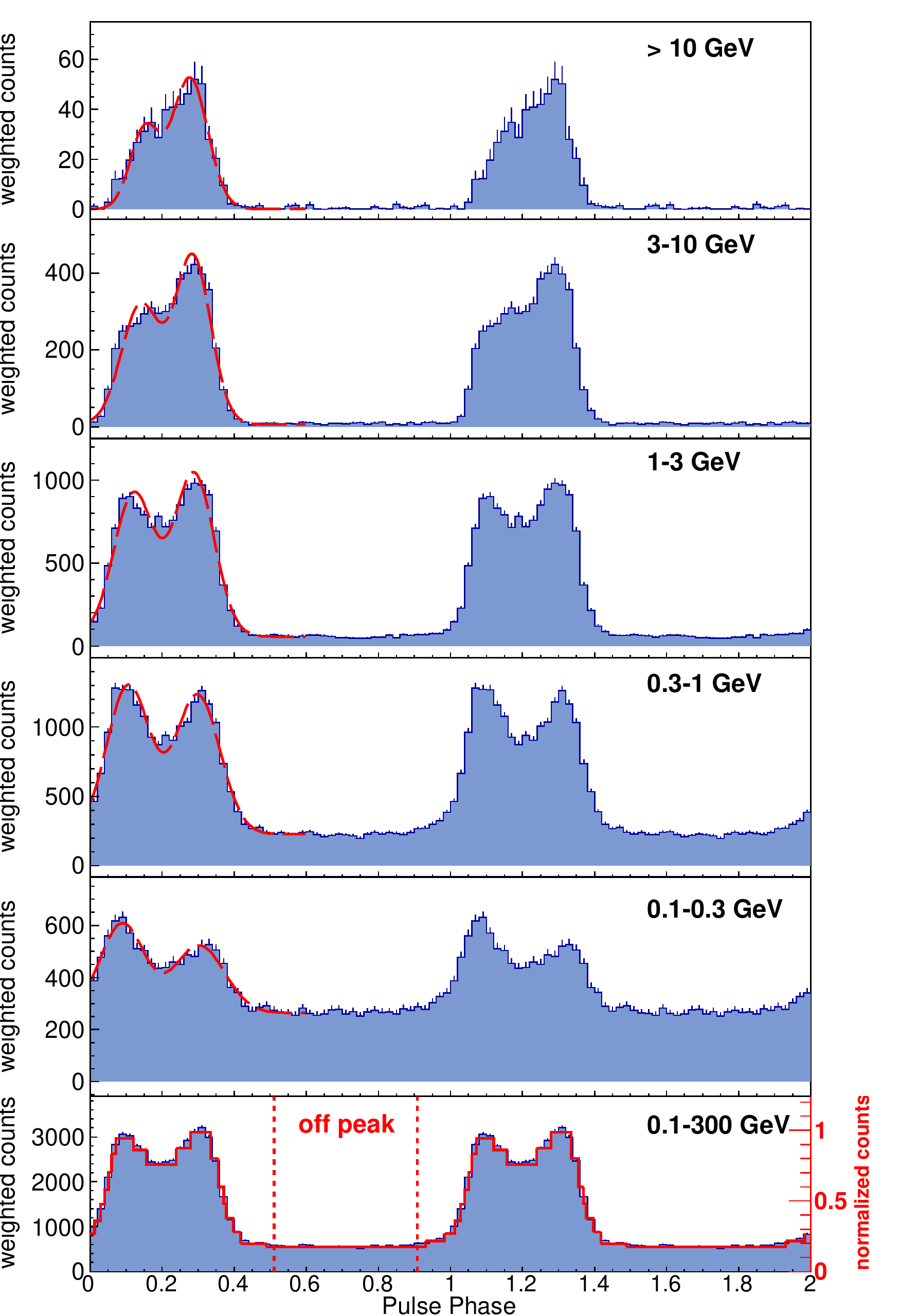}
\caption{Weighted pulse profile of \psrj\/ at different energies.
Two rotational pulse periods are shown, with a resolution of 50 phase bins per period.
The double Gaussian profile fitted to the light curves {is} shown with dashed red lines.
The bottom panel shows the weighted pulse profile above 100 MeV.
The Bayesian block decomposition is represented by red lines in the bottom panel.
The region indicated by the red dashed lines is the off-peak phase.}
\label{profile}
\end{figure*}
\end{center}

We divided the {pulse phase} of \psrj\/ into two parts, off-peak and on-peak intervals.
We begin by deconstructing the pulsed light curve into simple Bayesian Blocks using the same algorithm described in the 2PC, by Jackson et al. (2005) and Scargle et al. (2013).
To produce Bayesian Blocks on the pulsation light curve, we extended the data over three rotations, by copying and shifting the observed phases to cover the phase range from $-$1 to 2.
We define the final blocks to be between phases 0 and 1.
The lowest Bayesian Block is defined as the off-peak phase.
To avoid potential contamination from the trailing and/or leading edge of the peaks, we reduce the extent of the block by 10\% on either side, referenced to the center of the block.
The off-peak phase is located between $\phi$=0.511$-$0.909, {and is consistent} {with} the off-peak {definition for} \psrj\/ in the  2PC.
The on-peak phase is located at $\phi$=0.0$-$0.511 and  $\phi$=0.909$-$1.0.
In Figure \ref{profile},  the bottom panel shows the Bayesian block decomposition and the off-peak phase range.
{More discussion of Figure \ref{profile} is presented in Section \ref{light}.
}

\section{Off-peak analysis}

\subsection{Discovery and analysis of Fermi J0020+7328}

\begin{center}
\begin{figure*}
\centering
\includegraphics[scale=0.28]{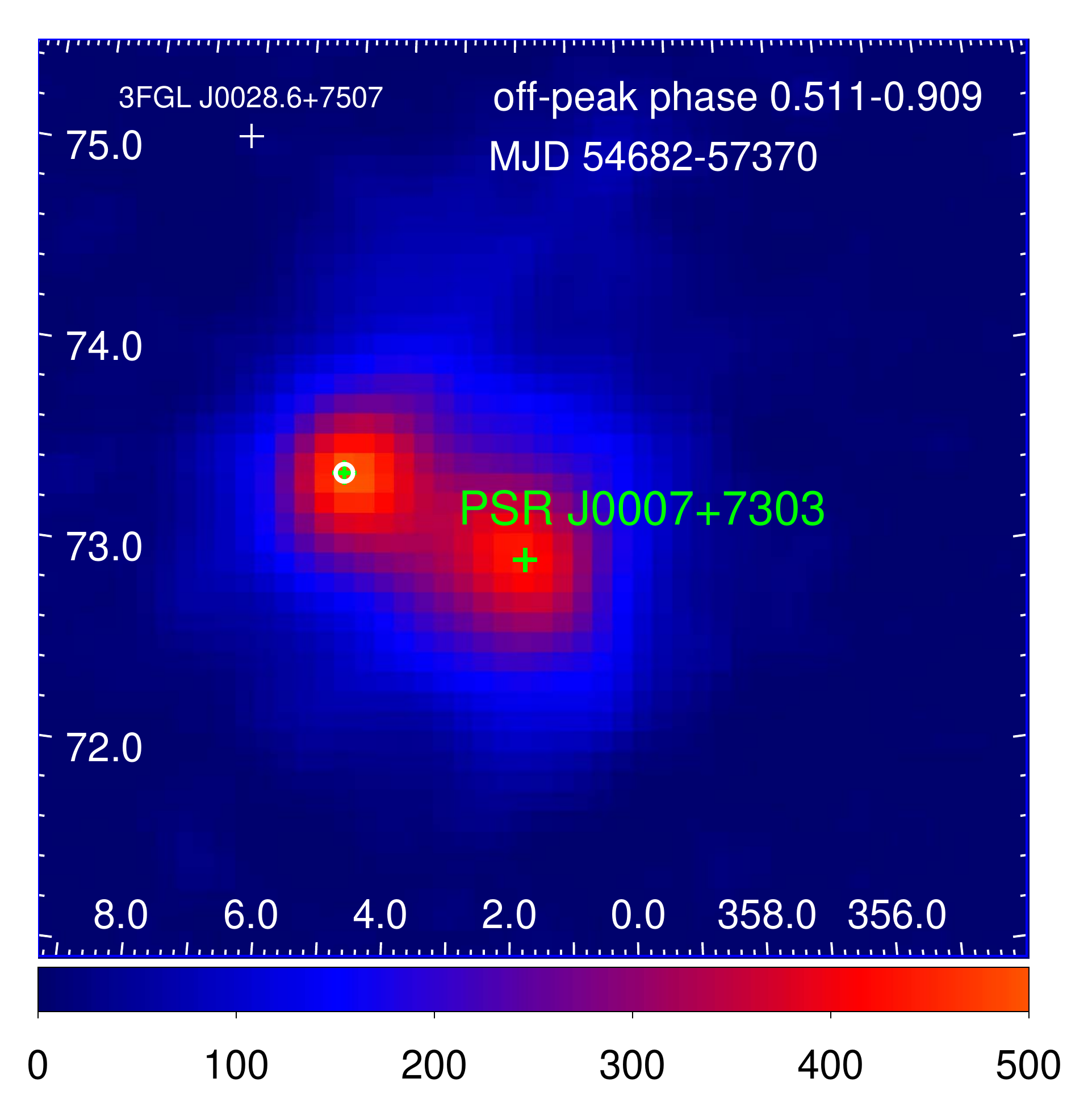}
\includegraphics[scale=0.28]{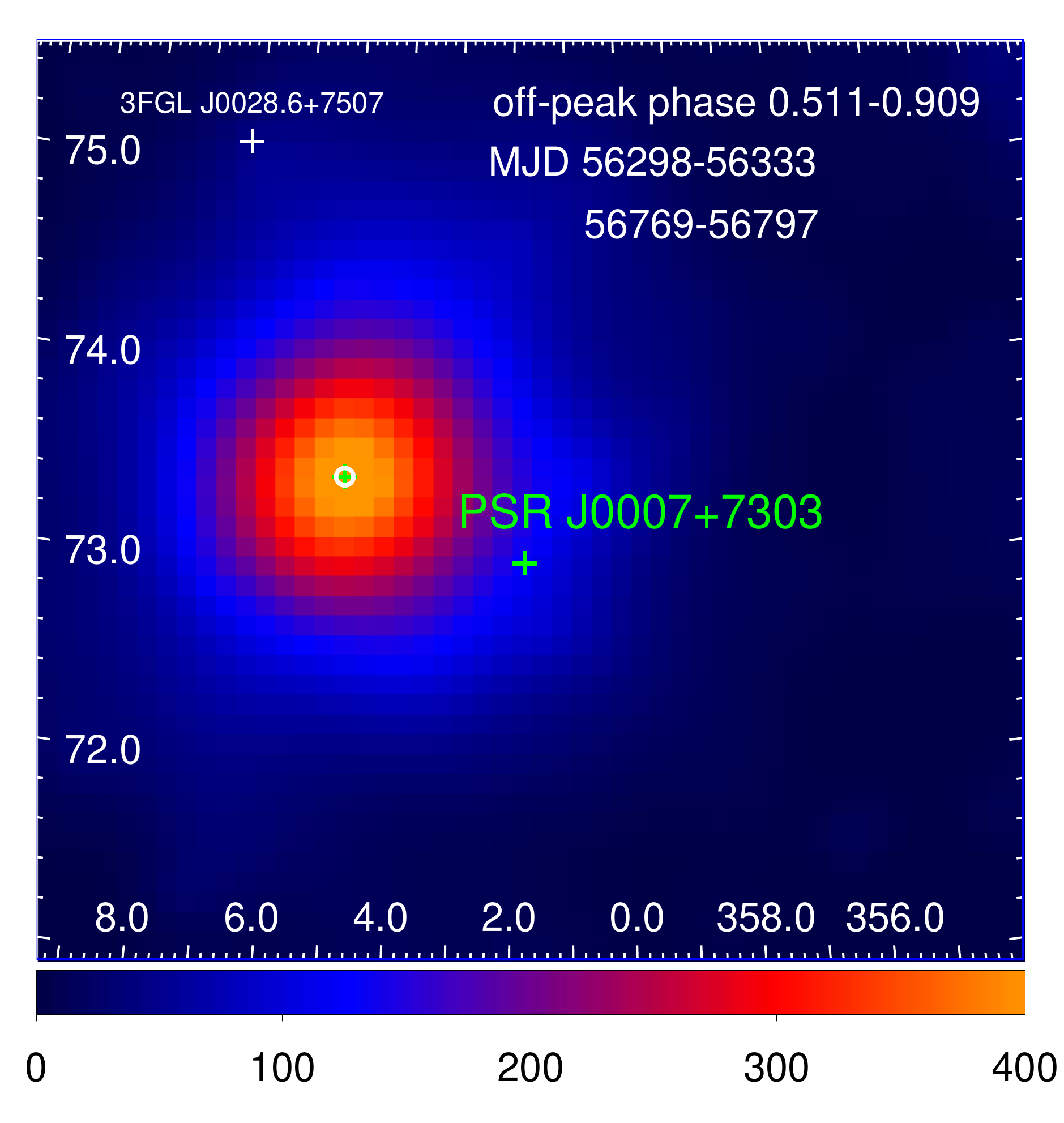}
\includegraphics[scale=0.28]{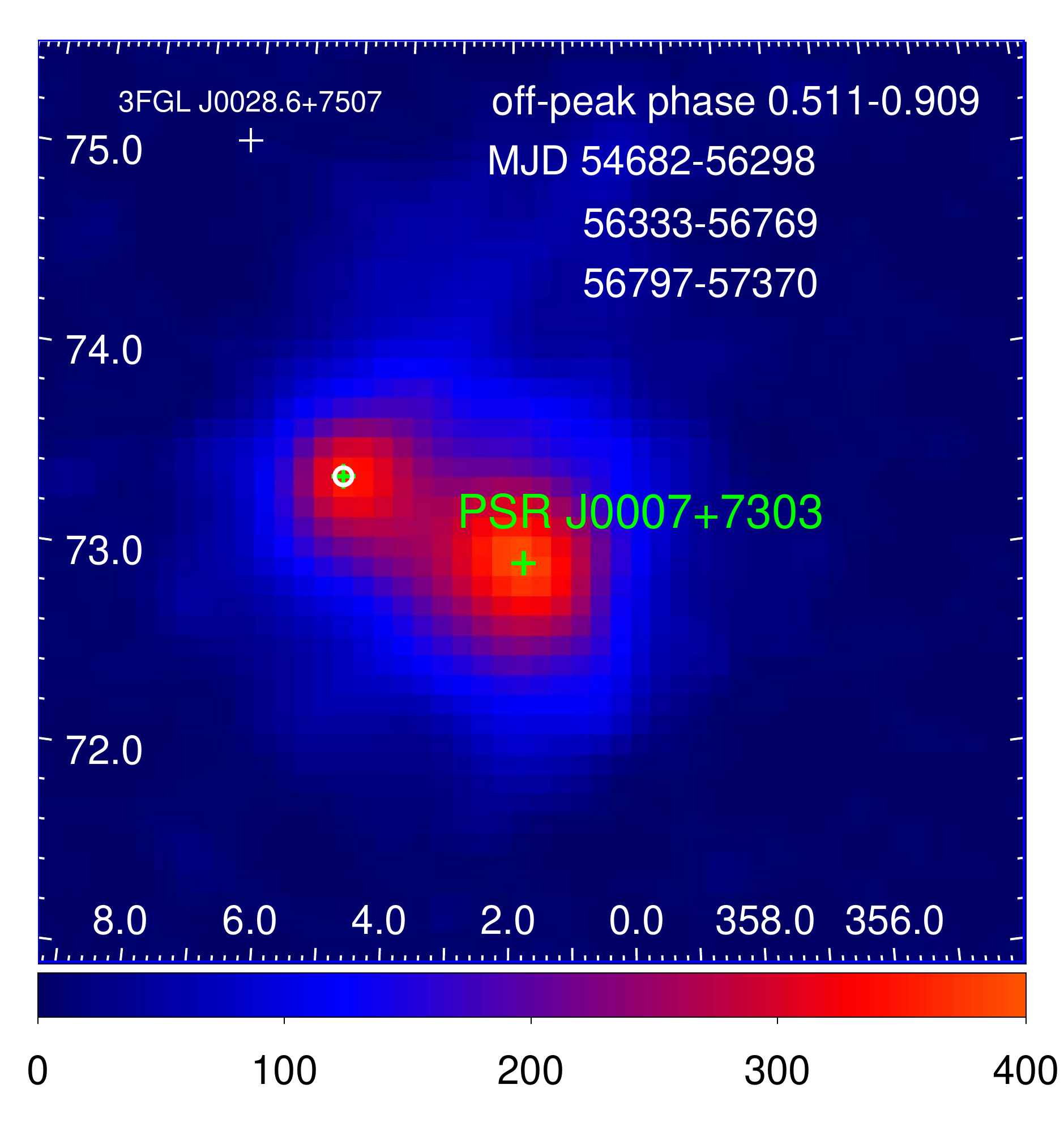}
\caption{Left: TS maps of the \fermi-LAT field surrounding \psrj\/ during the off-peak phase.
The fitted position and 95\% confidence error circle of a previously unknown gamma-ray source, Fermi J0020+7328, are shown with a green cross and a white circle.
{The X-and Y-axis are RA and DEC referenced at J2000.}
Middle: the same as in the left panel, but only integrating the {flare periods} of Fermi J0020+7328.
Right:  the same as in the left panel, but integrating only the off-flare periods of Fermi J0020+7328.
The off-peak phases and MJD ranges are shown in each panel.
See text for details. }
\label{ts_offpeak}
\end{figure*}
\end{center}

Figure \ref{ts_offpeak} (left panel) shows a TS map of the off-peak phase of \psrj\/.
In the vicinity of \psrj\/ there is a previously unknown gamma-ray source.
Applying \textit{Pointlike}, the best-fit position of this gamma-ray source above 100 MeV is {RA$_{J2000}$ =4$\fdg$973 and Dec$_{J2000}$ =73$\fdg$462}, with a 95\% confidence error circle radius of 0$\fdg$044.
By using the fitted position and assuming a power law spectral shape ($dN/dE=N_{0}(E/E_{0})^{-\Gamma}$ cm$^{-2}$ s$^{-1}$ GeV$^{-1}$),
the \textit{gtlike} analysis of this gamma-ray source resulted in a TS value of 315 (we shall refer to this source as Fermi J0020+7328 hereafter).
The TS value of \psrj\ is 281 in the off-peak phases.

\begin{center}
\begin{figure*}
\centering
\includegraphics[scale=0.5]{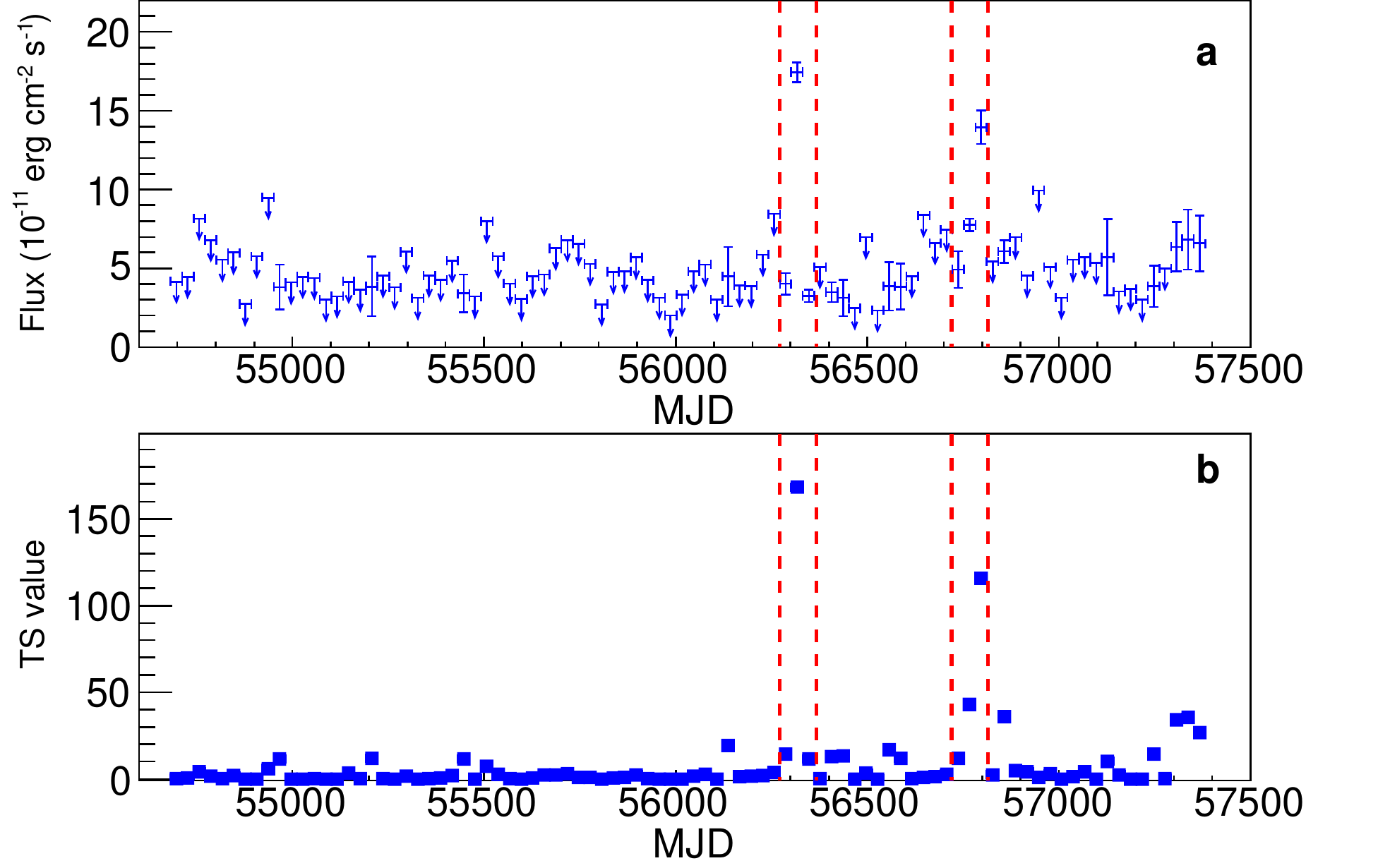}
\includegraphics[scale=0.5]{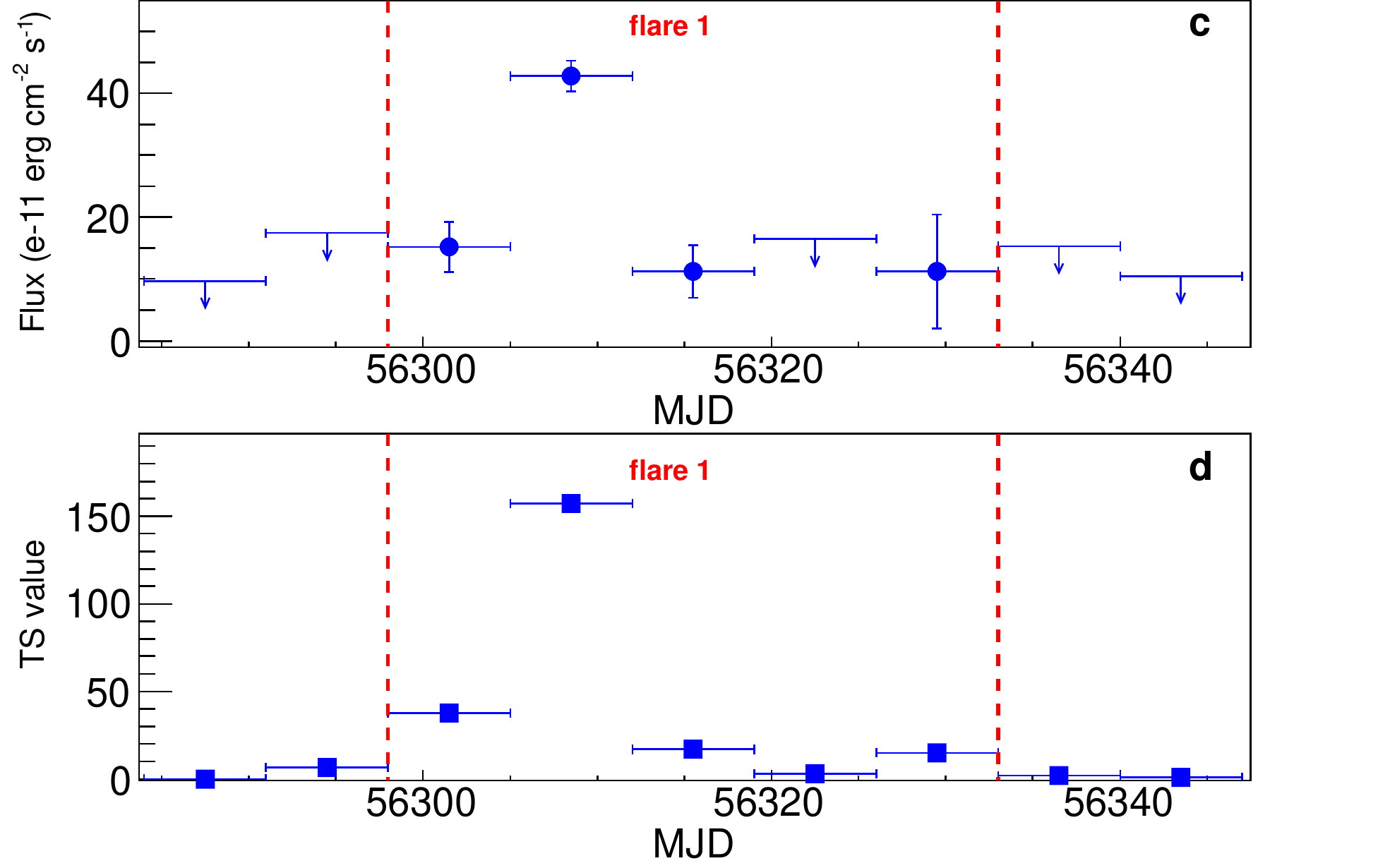}
\includegraphics[scale=0.5]{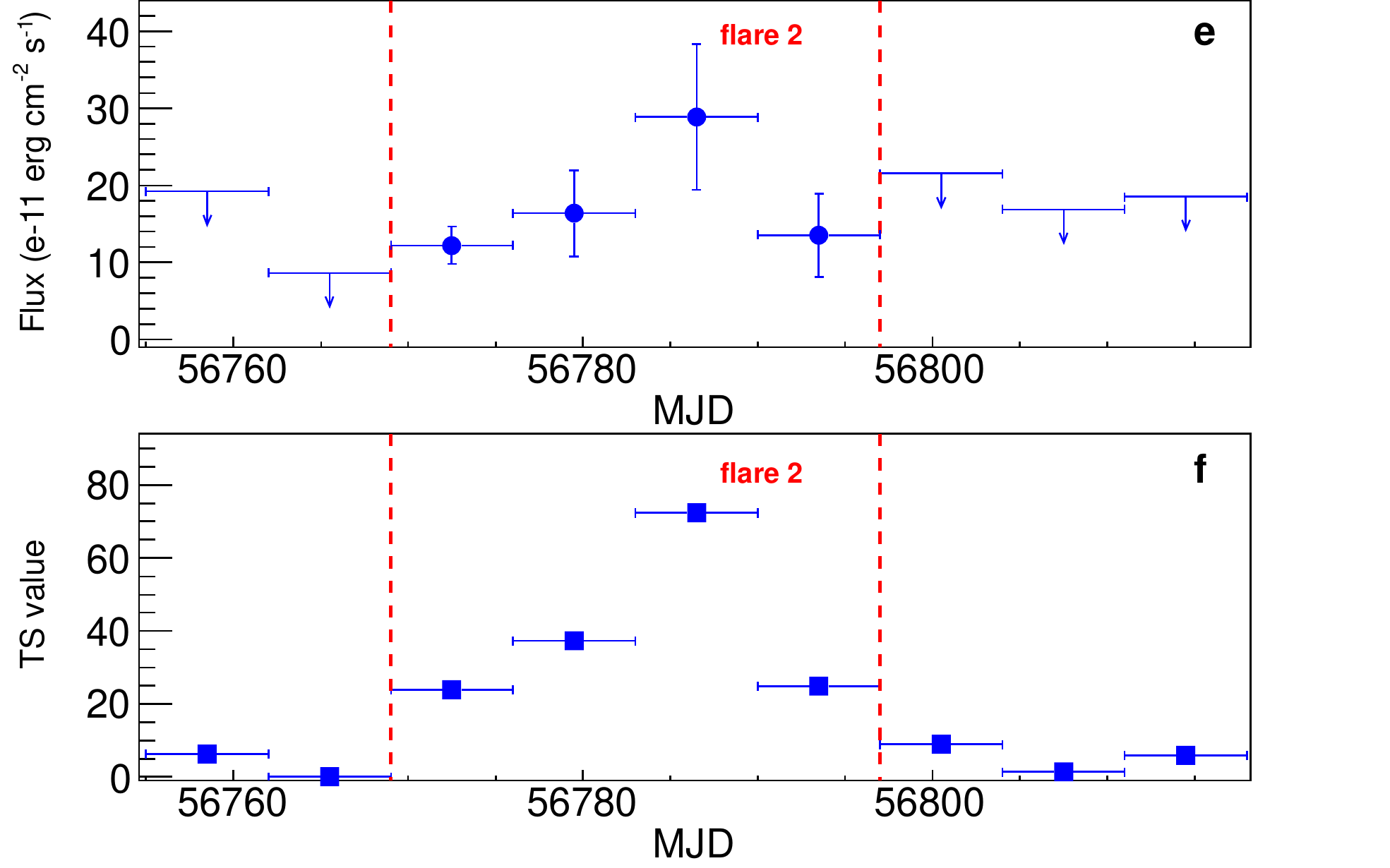}
\caption{Panels a \& b: Monthly gamma-ray flux and TS value of Fermi J0020+7328.
Two possible flares are indicated with red dashed lines.
Panels c \& d: Expanded scale, weekly gamma-ray flux and TS value evolution around flare 1.
Panels e \& f: Expanded scale, weekly gamma-ray flux and TS value evolution around flare 2.
The {periods} of the two flares defined in the text are indicated with red dashed lines in panels c -- f.}
\label{new_lc}
\end{figure*}
\end{center}

Assuming a power law spectral shape, we produced the monthly binned {long-term} light curve of Fermi J0020+7328  (Figure \ref{new_lc}, panel a \& b).
The 95$\%$ flux upper limits are calculated with Helene's method (Helene 1983) assuming a photon index of $2.0$ if the TS value is below 9.
Besides occasional flux fluctuations, two large flares are apparent, each lasting more than a month, which are indicated in Figure \ref{new_lc}, panels a and b. %
A weekly light curve with an expanded scale of the two flares are shown in Figure \ref{new_lc}, panel c -- f.
The flaring periods are MJD 56298 -- 56333 (flare 1) and MJD 56769 -- 56797 (flare 2), both of which are indicated in Figure \ref{new_lc}, panel c -- f.
TS maps of the \psrj\ region during flaring and non-flaring periods are shown in Figure \ref{ts_offpeak}.

\begin{center}
\begin{figure*}
\centering
\includegraphics[scale=0.4]{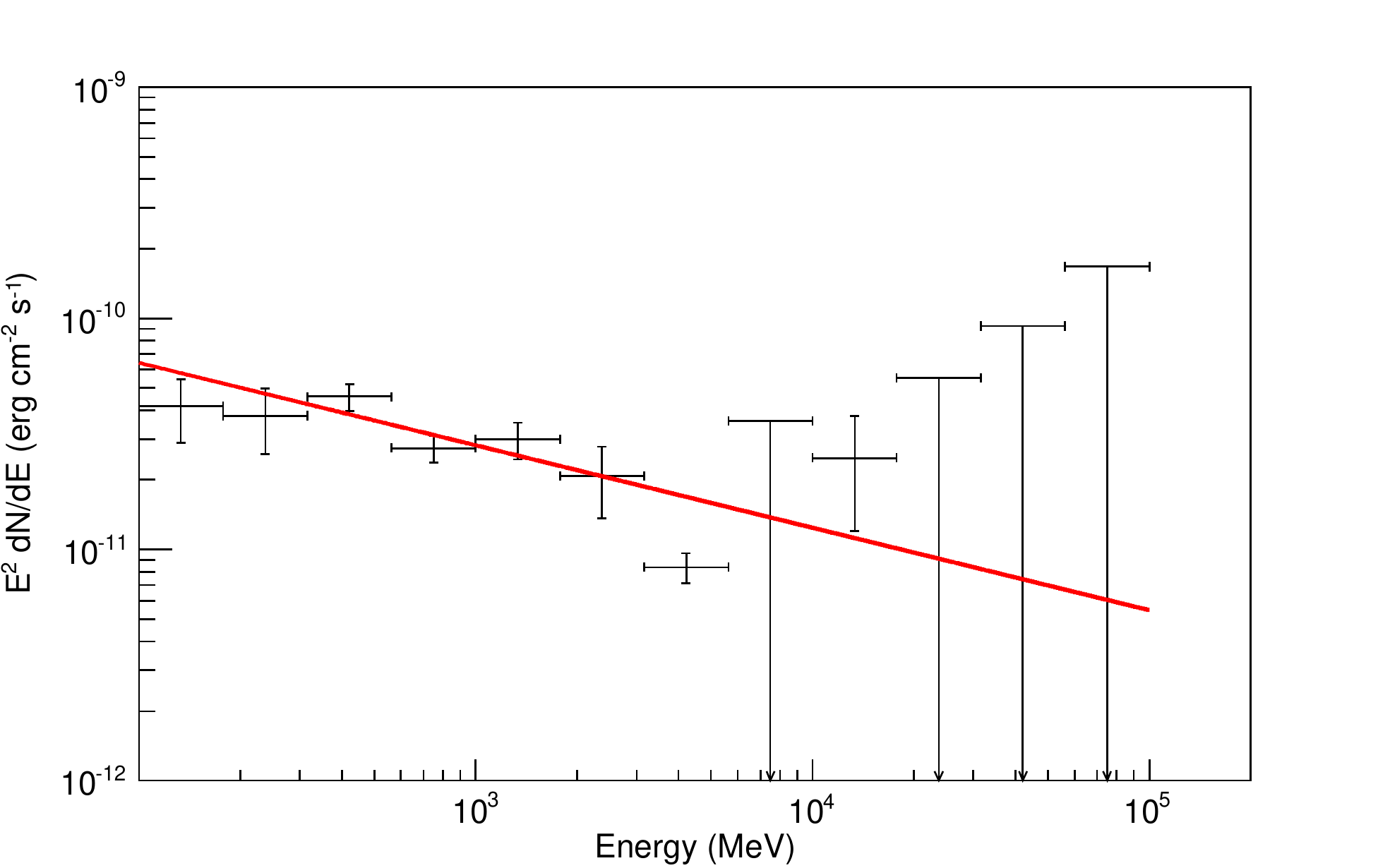}
\includegraphics[scale=0.4]{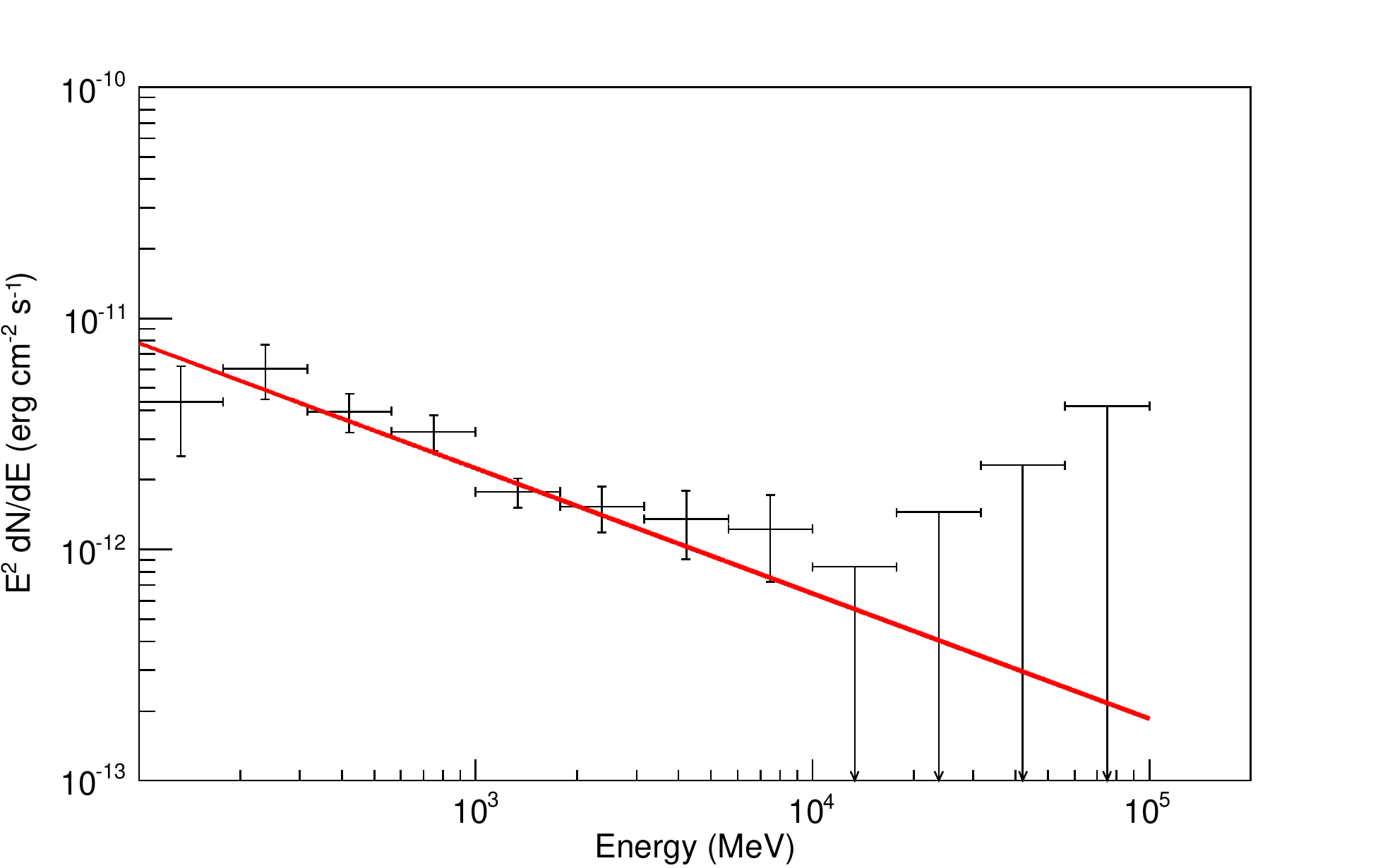}
\caption{{\em Fermi}-LAT spectra of the gamma-ray source Fermi J0020+7328 during flare (left) and non-flare (right) periods.
The \textit{gtlike} fitted models are shown with red lines.}
\label{new_sed}
\end{figure*}
\end{center}

The spectra of Fermi J0020+7328 in the different periods, both during the flares and in the off-flare period, are modeled by a
power law and a power law with an exponential cutoff ($dN/dE=N_{0}(E/E_{0})^{-\Gamma}$exp$(-E/E_{0}) $ cm$^{-2}$ s$^{-1}$ GeV$^{-1}$).
We compare the two models using the likelihood ratio test (Mattox et al. 1996).
The $\Delta$TS between the two models is less than 9, {which indicates that a cutoff is not significantly preferred}.
The best-fit spectral parameters and corresponding TS values are listed in Table \ref{new_fit}, while the spectral energy distributions (SEDs) are shown in Figure \ref{new_sed}.
The gamma-ray flux is $\sim$ 12 times higher and the spectrum is harder during the flare period than outside it.
We investigated the two flares individually.
The flux levels of {flares} 1 and 2 are consistent within errors while the spectrum of flare 1 is softer than that of flare 2 (Table  \ref{new_fit}).

Active galactic nuclei (AGNs) are the dominant source population of the GeV sky (Ackermann et al. 2015).
Since the gamma-ray flux of Fermi J0020+7328 is variable, it {displays} flares, and its spectrum is consistent with {those of} gamma-ray-detected AGNs (Ackermann et al. 2015), it is possible that this gamma-ray source is indeed an AGN.

\begin{center}
\begin{figure*}
\centering
\includegraphics[scale=0.38]{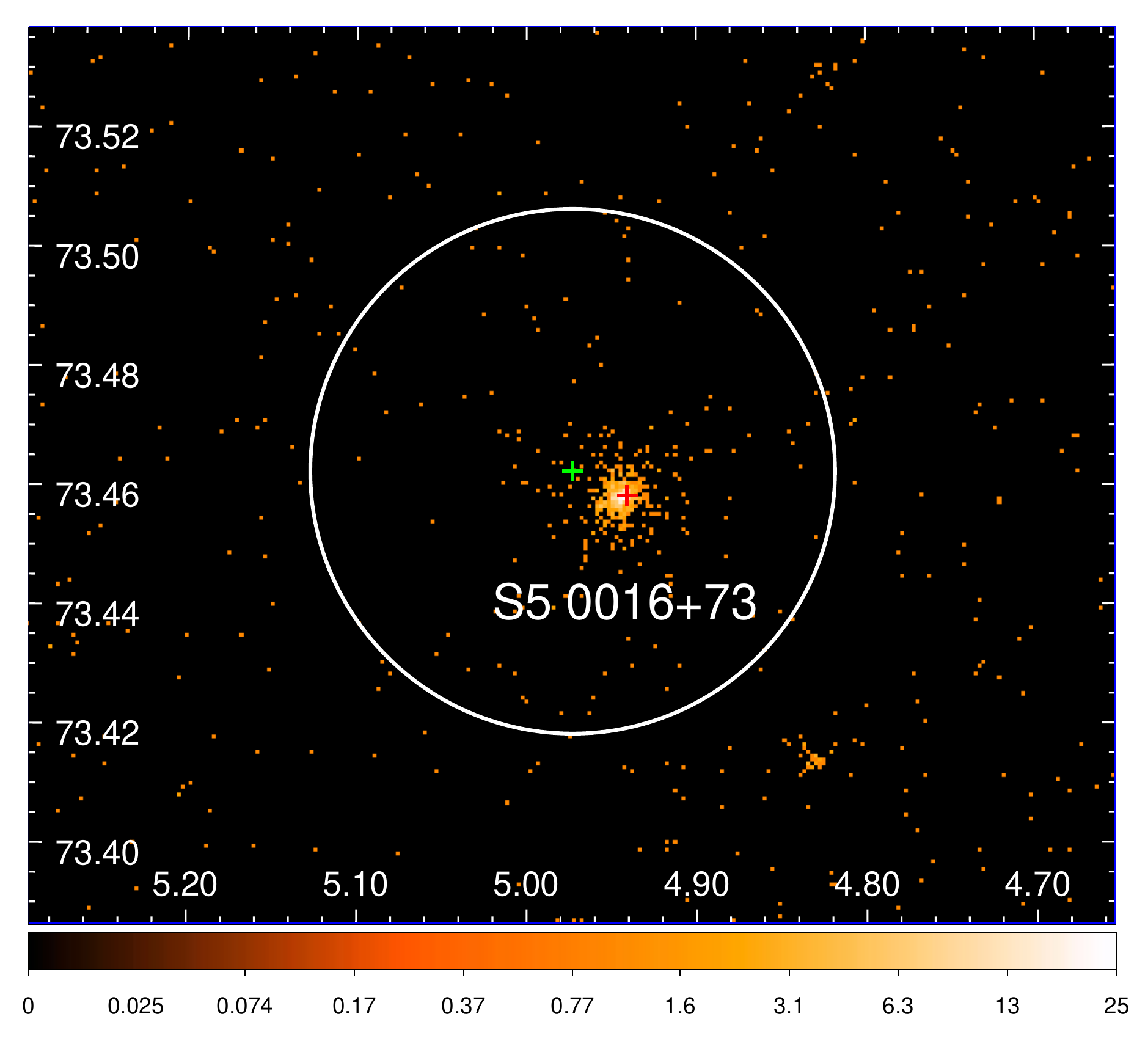}
\includegraphics[scale=0.45]{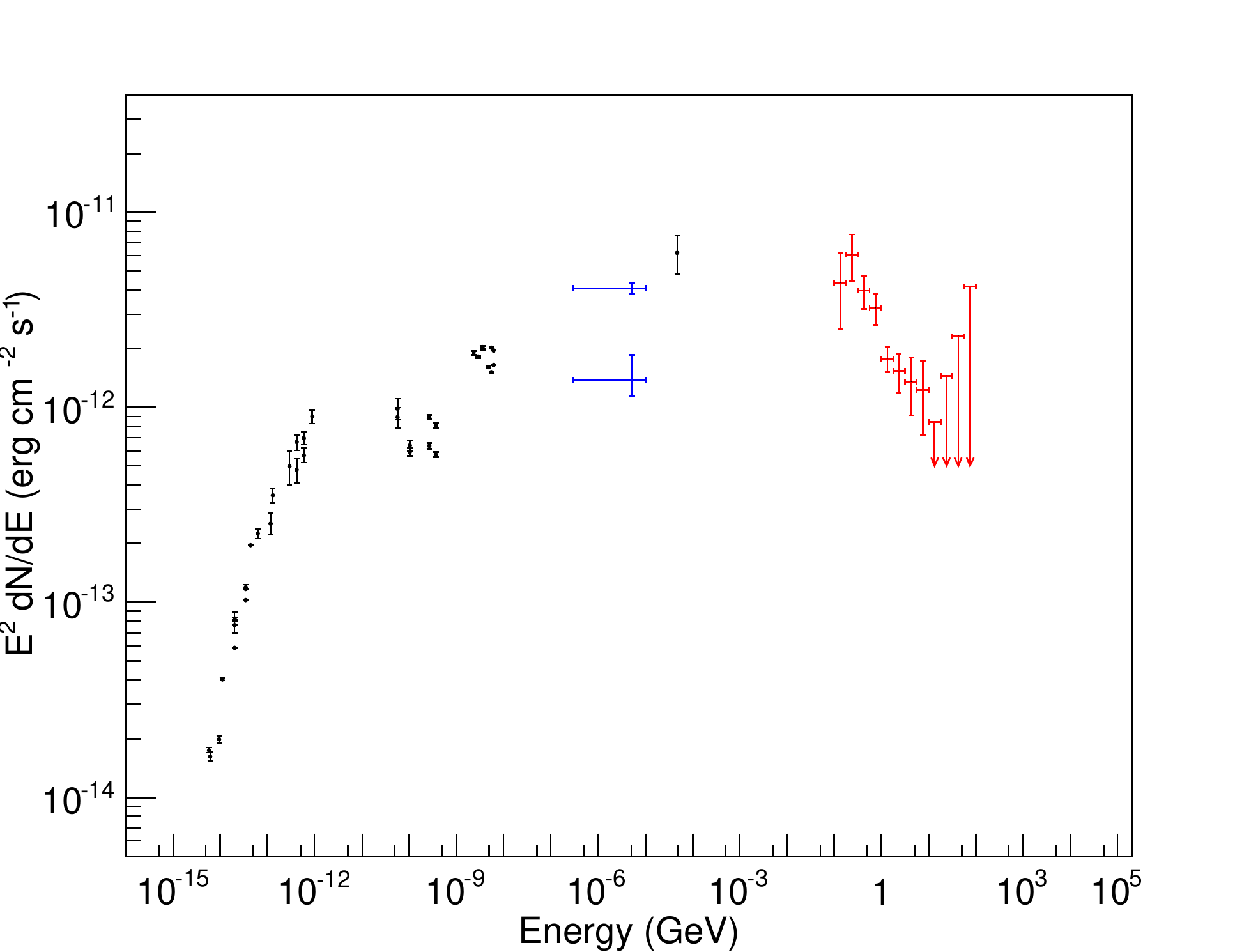}
\caption{Left: \emph{Swift}/XRT counts map from 2006 November 4 observation of the previously unknown gamma-ray source, Fermi J0020+7328, in the 0.3-10 keV range.
The \fermi-LAT best-fit position and 95\% confidence error circle are shown with a green cross and a white circle respectively.
The position of \qso\ is indicated with a red cross. The X- and Y-axes are RA and DEC referenced at J2000.
{Right: Non-simultaneous SED of \qso\/. The \emph{Swift}/XRT data are shown in blue while the off-flare SED of Fermi J0020+7328 is shown in red.
}}
\label{swift}
\end{figure*}
\end{center}

Between 2006 -- 2016 there have been seven observations of Fermi J0020+7328 with \emph{Swift}.
However, only two observations, 2006 November 4 (observation ID 00036187001) and 2009 September 11 (observation ID 00036187005), have sufficient \emph{Swift}/XRT exposure (9.6 ks and 7.4 ks, respectively) for {spectral} analysis.
The \emph{Swift}/XRT counts map is shown in Figure \ref{swift}, left panel.
{The quasar \qso\ is the only X-ray source detected within the error circle of gamma-ray source Fermi J0020+7328 and is only 0$\fdg$01 away, which argues for a possible association.
\qso\/ is a Flat Spectrum Radio Quasar (FSRQ) with redshift of 1.781 (Lawrence et al. 1986).
The \emph{Swift}/XRT X-ray spectra of \qso\ is well fit with a power law and the spectral parameters are shown in Table \ref{swift_fit}.
The X-ray photon index measured by \emph{Swift} is consistent with a previous \emph{ROSAT} result (Donato et al. 2001).
Between the two \emph{Swift}/XRT observations, there is also significant X-ray flux variability.
However, because of the large uncertainty of the spectral index for the 2009 September 11 (observation ID 00036187005), we cannot claim a spectral change.
With archival multi-wavelength data collected using the ASDC online services\footnote{\url{http://tools.asdc.asi.it/SED/}}, we show the SED of \qso\/ including \emph{Swift}/XRT data and the off-flare SED of Fermi J0020+7328 measured by \emph{Fermi}-LAT in Figure \ref{swift}, right panel.
The off-flare SED of Fermi J0020+7328 is consistent with the overall SED of \qso\/.
We propose that Fermi J0020+7328 is the GeV counterpart of \qso\/.
Its gamma-ray photon index and flux level during the off-flare period are at the average of {\em Fermi}-LAT detected FSRQs (Ackermann et al. 2015).
The flux level of \qso\/ during its flare period is in the upper end of gamma-ray-detected FSRQs.
Considering its distance, the gamma-ray luminosity of \qso\/ is common among {\em Fermi}-LAT detected FSRQs.
}

\qso\ is only $\sim$1 degree away from \psrj, and {is more significant than the pulsar during off-peak phases} (Figure \ref{ts_offpeak}, left panel).
Taking the proximity and the size of the \emph{Fermi}-LAT PSF into consideration, \qso\ may affect the results obtained from \psrj\/.
To minimize its influence, we carried out the {\em Fermi}-LAT analysis of \psrj\ during the non-flare period of \qso\ for both the off-peak and on-peak phases of \psrj\/.

\begin{table*}{}
\centering
\scriptsize

\caption{{\em Fermi}-LAT fitted spectral parameters of Fermi J0020+7328 during the off-peak phase of \psrj.}

\begin{tabular}{llll}
\hline
\hline
\\
Time Interval & Spectral Index & TS  & Energy Flux, 0.1--300 GeV  \\
                &                  &                     &10$^{-11}$ erg~cm$^{-2}$s$^{-1}$                                  \\
\\
\hline\hline                 
\\

Flaring intervals combined         &  2.36 $\pm$ 0.01 $\pm$ 0.04       &  357    & 16.7 $\pm$ 0.8 $\pm$ 0.3 \\
Flare 1 (MJD 56298--56333)      &  2.43 $\pm$ 0.02 $\pm$ 0.03   &  197    &  16.4 $\pm$ 0.8  $\pm$ 0.5 \\
Flare 2 (MJD 56769--56797)     &   2.22  $\pm$ 0.12 $\pm$ 0.07   &  134   & 16.6 $\pm$  3.2 $\pm$1.5 \\
\\
\hline\
\\
Non-flaring intervals         &  2.54 $\pm$ 0.06 $\pm$ 0.16     &201  &1.4 $\pm$ 0.2 $\pm$ 0.1 \\
\\
\hline\hline                 
\label{new_fit}
\tablecomments{The first (second) uncertainties correspond to statistical (systematic) errors.}

\end{tabular}
\end{table*}

\begin{table*}{}
\centering
\scriptsize
\caption{\emph{Swift}/XRT fitted spectral parameters of \qso.}
\begin{tabular}{cccccc}
\\
\hline\hline                 
\\
Observation ID & Date & Spectral Index & Flux, 0.3--10 keV &    N$_{H}$    &Reduced $\chi^{2}$ / (D.O.F.)\\
                &                  &                     &10$^{-12}$ erg~cm$^{-2}$s$^{-1}$ & 10$^{22}$ cm$^{-2}$ &                              \\

\\
\hline\hline                 
\\
00036187001        &  2006 November 4   &  1.51$\pm$0.16    & 4.1$\pm$ 0.3  & 0.18$\pm$0.07 & 0.59 (21)  \\
00036187005       & 2009 September 11  &   1.81$_{-0.41}^{+0.44}$  & 1.4 $_{-0.2}^{+0.5}$ & 0.31$_{-0.27}^{+0.30}$& 1.38 (3)\\
\\
\hline\hline                 
\\
\label{swift_fit}
\end{tabular}
\end{table*}

\subsection{\psrj\/}

\begin{center}
\begin{figure*}
\centering
\includegraphics[scale=0.4]{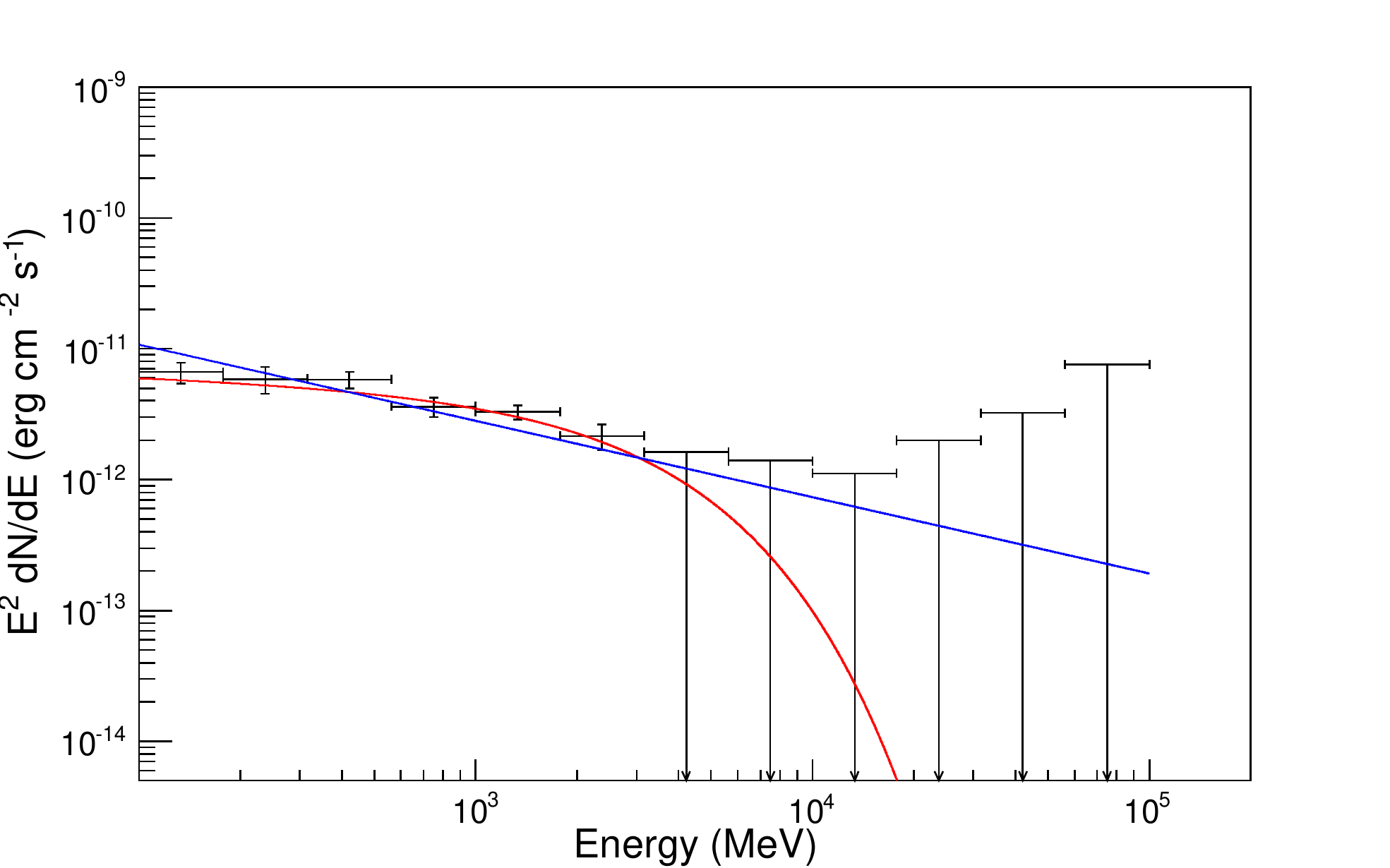}
\includegraphics[scale=0.4]{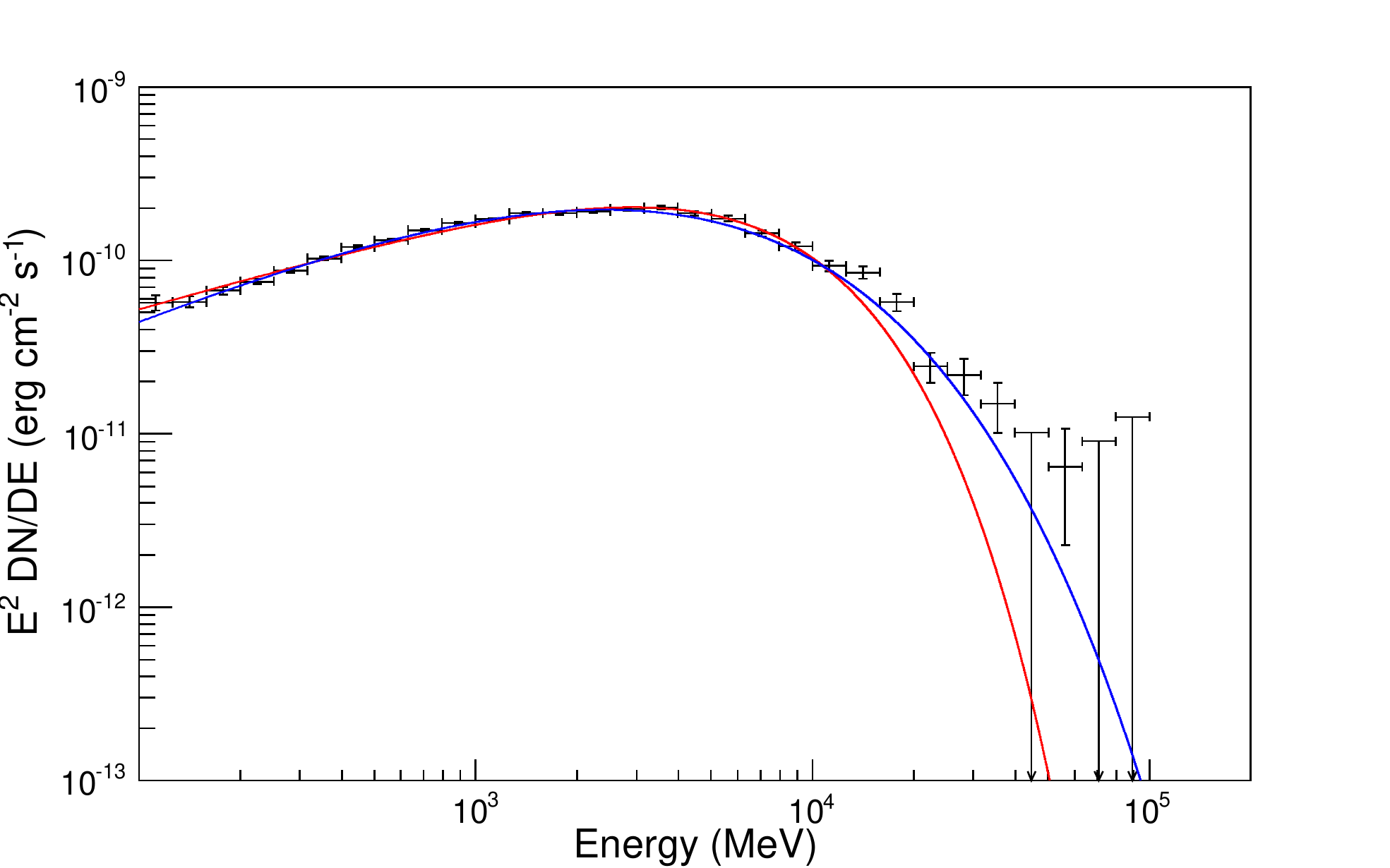}
\caption{{\em Fermi}-LAT spectra of \psrj\ during off-peak (left) and on-peak (right) phases.
Maximum likelihood models fitted with gtlike are shown with red lines (power law with exponential cutoff) and blue lines (power law, left and power law with sub-exponential cutoff, right).}
\label{psrj_sed}
\end{figure*}
\end{center}

As mentioned, we carried out the off-peak analysis of \psrj\ during the non-flare period of \qso\/.
The off-peak emission of \psrj\ was reported previously by Abdo et al. (2012) using two years of data with P6V11 IRFs and Abdo et al. (2013) using three years of data with P7V6 IRFs (P6V11 and P7V6 are both previous versions of LAT IRFs\footnote{\url{http://fermi.gsfc.nasa.gov/ssc/data/analysis/documentation/Cicerone/Cicerone\_LAT\_IRFs/IRF\_overview.html}}), yielding TS values of 40 and 71.5, respectively.
However, no spectral cutoff was detected in these analyses, probably due to the limited statistics.
To explore the possible cutoff in the {off-peak spectrum}, we modeled \psrj\ using a power law function with and without an exponential cutoff.
We compare the two models using the likelihood ratio test, leading to a $\Delta$TS of 12.4, which indicates that the significance of the spectral cutoff is $\sim$3.5 $\sigma$.
{\psrj\ is detected with a TS value of 262} and the fitted parameters are shown in Table \ref{psrj_fit}.
The SED is shown in Figure~\ref{psrj_sed}.
This is the first hint of the existence of a spectral cutoff during the off-peak phase of \psrj\/.

{Abdo et al. (2012) excluded a point-like hypotheses for the off-pulse emission at the 95\% level and thus reported a marginal detection of extent while in the 2PC the TS$_{ext}$ was calculated to be 10.8 ($\sim$3.3 $\sigma$).
With the use of more advanced response and diffuse models, significantly more data accumulated, {and the newly detected gamma-ray source Fermi J0020+7328 included in the model,} we performed the likelihood analysis again to check for the possible extension.
In the off-peak gamma-ray emission, we fitted an extended disk to \psrj\/ using \textit{Pointlike},
%
yielding a TS$_{ext}$ = 1.3; the disk is not favored.
The off-peak gamma-ray emission of \psrj\ is not extended.}

{As a check, adopting only the two years of data analyzed by Abdo et al. (2012), we fitted an extended disk to \psrj\/ during the off-peak phases using \textit{Pointlike} with Fermi J0020+7328 included in the model, leading to
%
TS$_{ext}$ = 2.3 (87\% confidence level), which is consistent with the 95\% confidence level measured by Abdo et al. (2012) and does not imply a significantly extended emission either.
{For a further check, we excluded Fermi J0020+7328 from the model and repeated the above analysis.
It leads to a disk of radius 0$\fdg$62$\pm$0$\fdg$13 located at RA$_{J2000}$ =1$\fdg$847 and Dec$_{J2000}$ = 73$\fdg$219 and a TS$_{ext}$ = 9.4 (99.9\% confidence level).
These values are consistent with the 0$\fdg$7$\pm$0$\fdg$3 extension at 95\% confidence level reported in Abdo et al. (2012).}
Thus, considering both the hint of the spectral cutoff at more than 3 $\sigma$ and the point-like morphology, we propose that the off-peak gamma-ray emission of \psrj\ originates from the magnetosphere of the pulsar rather than the PWN or the SNR CTA~1.
}

\begin{table*}{}
\centering
\scriptsize
\caption{{\em Fermi}-LAT spectral parameters of \psrj\ during off-peak and on-peak phases.}
\begin{tabular}{cccccc}
\hline
Phase Interval&Spectral Index& Cutoff Energy  & $b$ &TS&Flux, 0.1--300 GeV  \\
                &                  &                             (GeV)  &       &      &10$^{-11}$ erg~cm$^{-2}$s$^{-1}$\\

\\
\hline\hline                 
\\
off-peak         &  2.09$\pm$0.21 $\pm$ 0.83      &  2.7 $\pm$ 1.2 $\pm$ 1.3 & -   &262  &1.5$\pm$ 0.2 $\pm$ 0.1 \\

on-peak         &  1.44$\pm$ 0.01 $\pm$ 0.03   &   5.2  $\pm$ 0.1 $\pm$ 0.4 &  -    &190317   & 68.3 $\pm$0.5 $\pm$ 0.4 \\
                      &  1.13$\pm$ 0.06 $\pm$ 0.15      &   1.14  $\pm$ 0.36  $\pm$ 0.50  &  0.57 $\pm$0.04 $\pm$ 0.06  & 190278  & 68.1$\pm$0.5 $\pm$ 1.9  \\
\\
\hline\hline                 
\\

\label{psrj_fit}
\tablecomments{The first (second) uncertainties correspond to statistical (systematic) errors.}
\end{tabular}
\end{table*}


\section{On-peak analysis}
\label{onphase}


{For the on-peak analysis, the normalizations of all} 3FGL sources within three degrees of \psrj\ were left free in the model.
For Fermi J0020+7328 and all the 3FGL sources beyond three degrees, the {normalizations} adopted were that of the off-peak fitted values rescaled to the {ratio of the width of the on-peak to off-peak phase intervals and then fixed.
All other spectral parameters were fixed at off-peak fitted value except for \psrj\/.
In modeling the on-peak phase of \psrj\/, we have first considered a power law with an exponential cutoff.
The best-fit parameters are shown in Table {\ref{psrj_fit}}.

The SED of the on-peak phase is shown in Figure \ref{psrj_sed} and the fitted power law with an exponential cutoff spectral shape is shown with a red line.
At higher energies, the SED points deviate from the fitted spectral shape.
We studied alternative spectral shapes, specifically a power law with a sub-exponential cutoff ($dN/dE=N_{0}(E/E_{0})^{-\Gamma}\exp(-E/E_{0})^{b} $ cm$^{-2}$ s$^{-1}$ GeV$^{-1}$, leaving the exponential index $b$ free).
The best-fit parameters are shown in Table \ref{psrj_fit}.
The parameter $b$ is found to be 0.57$\pm$0.04$\pm$ 0.06 (the first/second uncertainties correspond to statistical/systematic errors.), which is consistent with the $b$ value for young pulsars derived by a stacking analysis (McCann 2015).
Using the likelihood ratio test, the $\Delta$TS between the two models of the region is 125, which indicates that the significance of the sub-exponential cutoff is $\sim$11 $\sigma$.
The fitted power law with sub-exponential cutoff spectral shape is shown with a blue line in Figure \ref{psrj_sed}, and it models the SED well.

As previously reported for the Vela and Geminga {pulsars} (Abdo et al. 2010, Bonnefoy et al. 2015), a value of $b<$1 could be interpreted as a blend of different phase-resolved $b$=1 spectra having different cutoff energies.
To explore this possibility, we carried out a phase-resolved spectral analysis.
The best-fit model from the on-peak phase-averaged fit is used as the input model for the phase-resolved fit: i.e.,  for this analysis, all parameters except those associated with the pulsar were fixed to the value derived from the on-peak phase fit.
In each phase range, the pulsar spectrum is modeled as both a power law with a simple exponential cutoff and with a sub-exponential cutoff.
Figure \ref{resolved} shows the phase-resolved spectral parameters of \psrj\/ in different phase bins.
From the phase-resolved fits, we note that the $b$ parameter is consistently lower than 1 in all tested phase bins.
We further note that for each phase bin, the significance of the sub-exponential cutoff is {greater than} 3 $\sigma$.
Thus, for the first time we have shown that both the on-peak averaged spectrum as well as the phase-resolved spectra of \psrj\ are better described
by a power law with a sub-exponential cutoff function.

\begin{center}
\begin{figure*}
\centering
\includegraphics[scale=0.4]{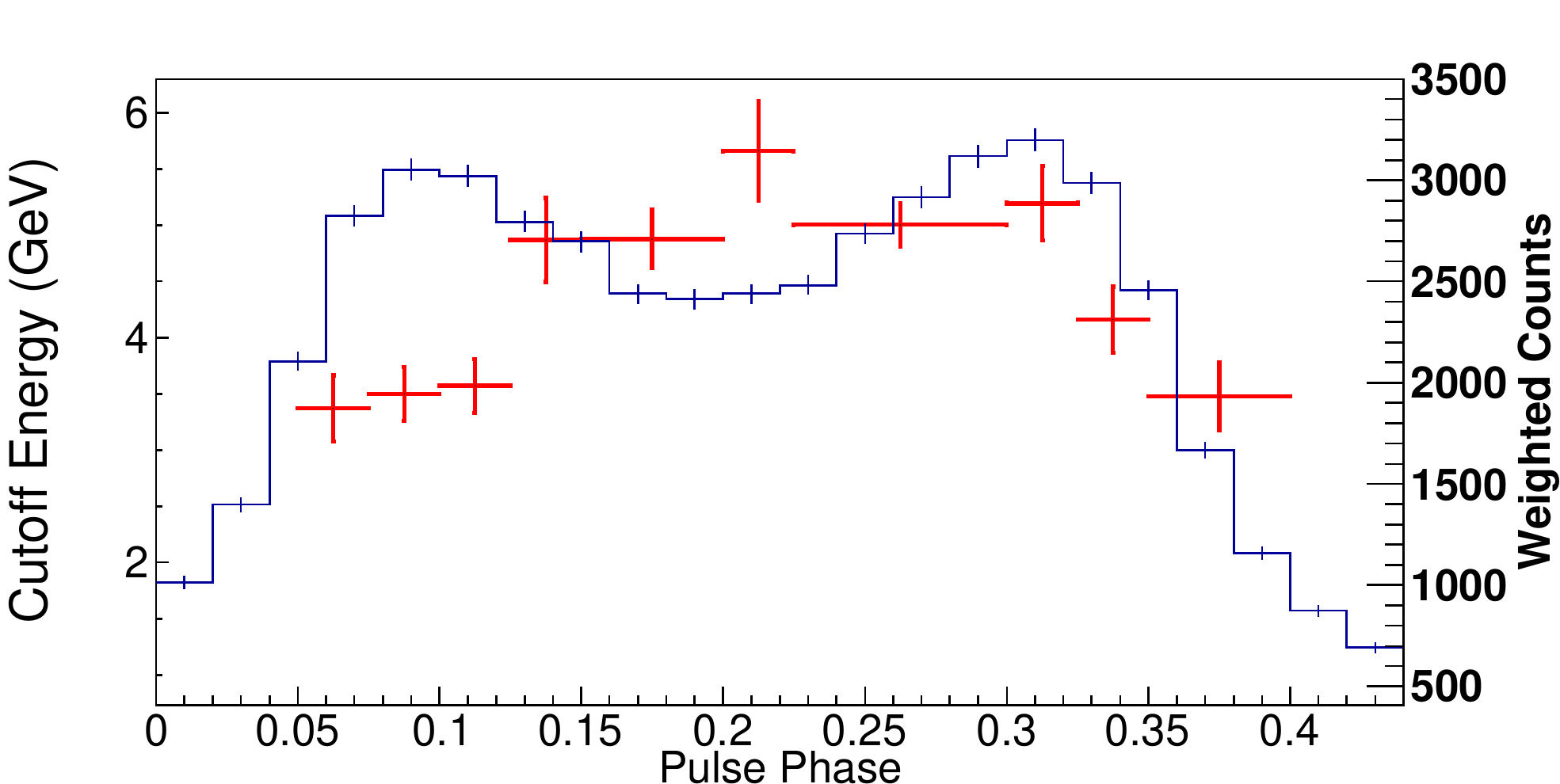}
\includegraphics[scale=0.4]{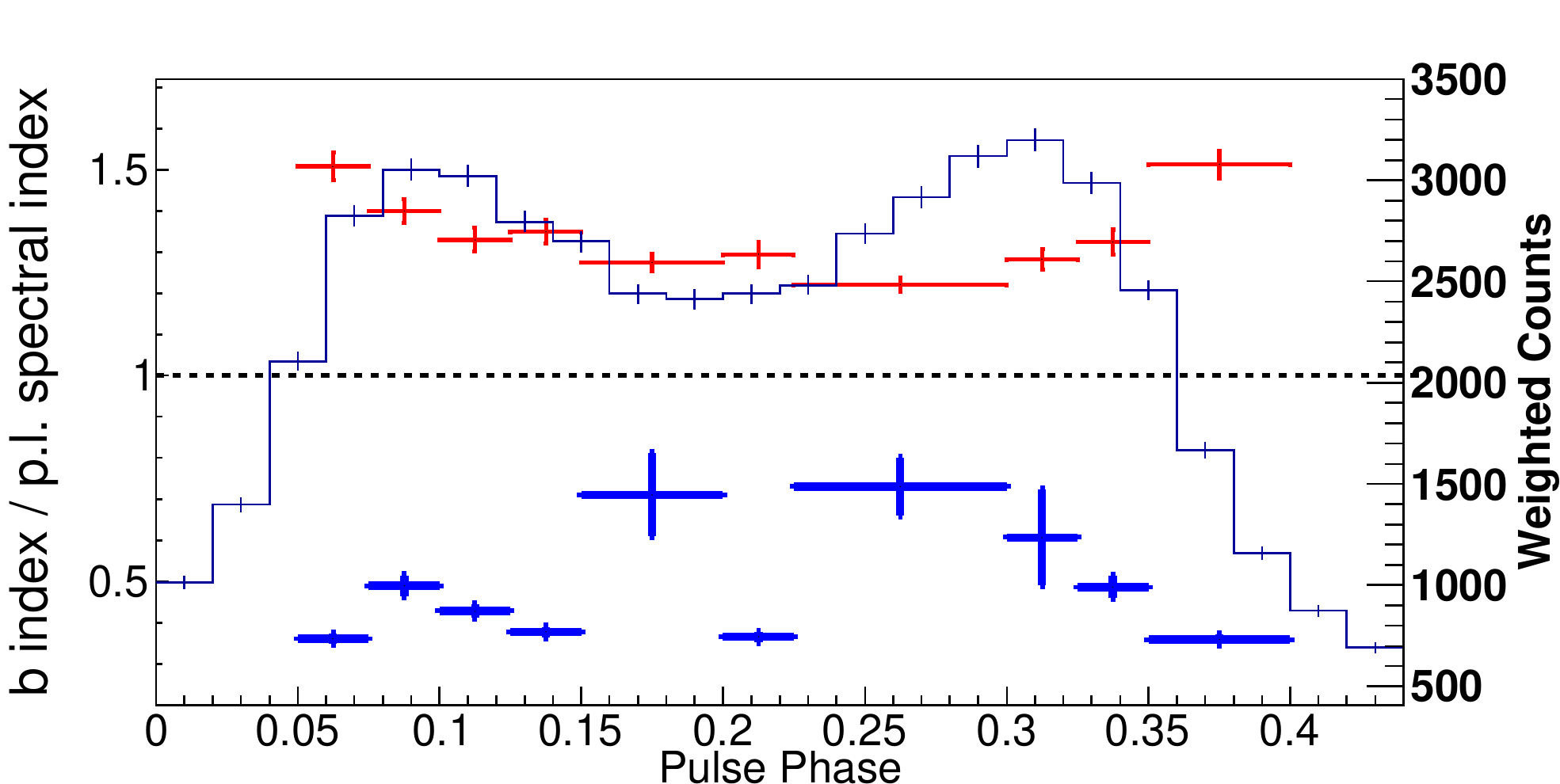}
\caption{The phase-resolved spectral parameters of \psrj\/.
The histogram in each panel shows the weighted phaseogram of \psrj\/ between 0.1 and 300 GeV (same as Figure \ref{profile}, bottom panel).
{The range in pulsar phase} is restricted to 0.0 -- 0.44.
Red points in the left and right panels correspond to the cutoff energy and the spectral index of the power law with exponential cutoff model, respectively.
The blue points in the right panel show the values of $b$ in the model of a power law with sub-exponential cutoff.
The dotted horizontal line indicates an index value of unity.}

\label{resolved}
\end{figure*}
\end{center}


\section{Light curve and spectral variability analysis}
\label{light}

\begin{center}
\begin{figure*}
\centering
\includegraphics[scale=0.7]{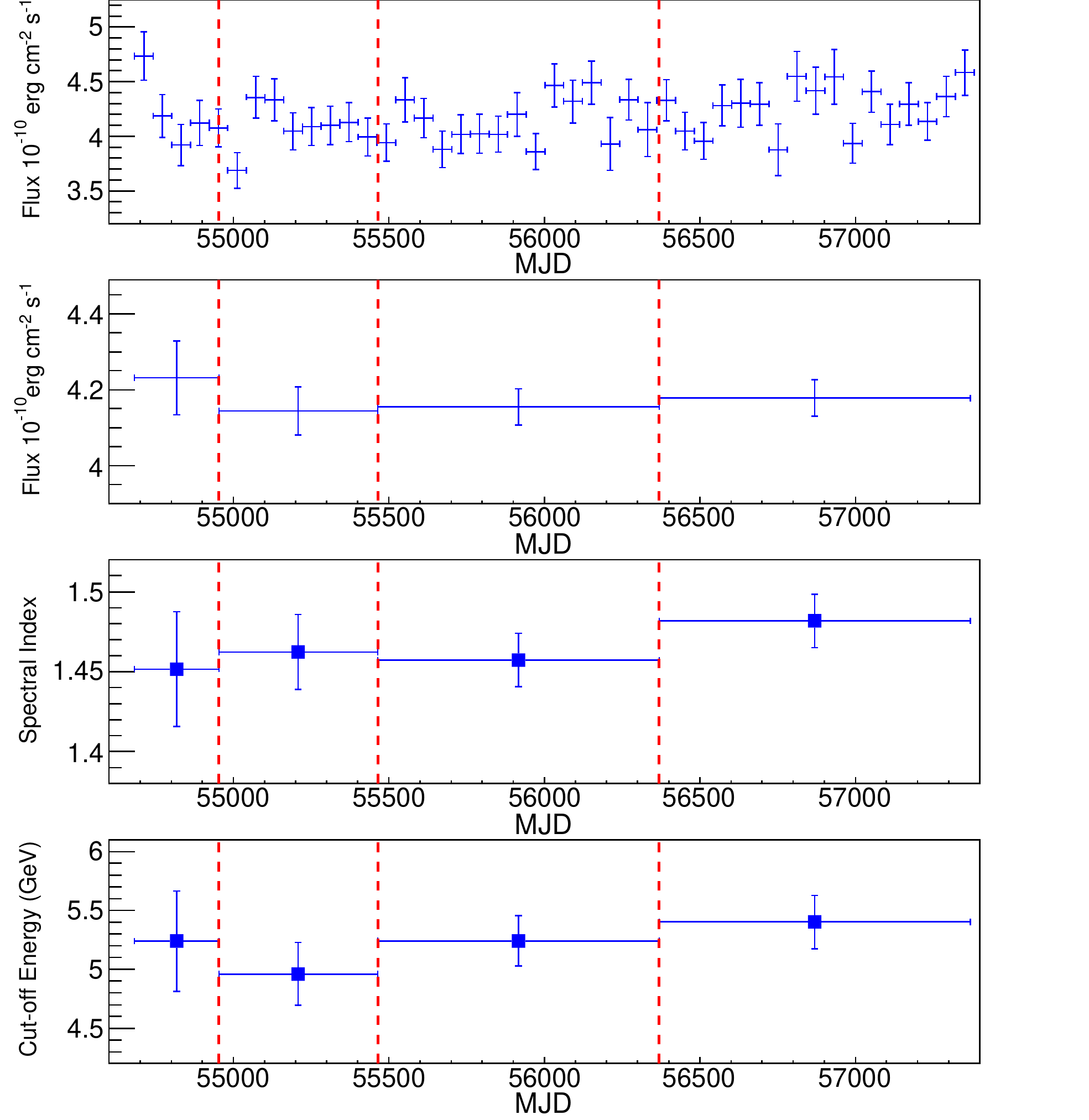}
\caption{From top to bottom: flux of \psrj\ in 0.1-300 GeV as a function of time in 60-day time bins; integrated flux, spectral index and cut-off energy of \psrj\ in 0.1-300 GeV for epochs separated by the glitches.
The dashed vertical red lines mark the time of glitches.}

\label{long_LC}
\end{figure*}
\end{center}

To check for long-term flux variability of \psrj\/ we performed a phase averaged likelihood analysis similar to that done in Section \ref{onphase}.
We selected the length of the time bin as 60 days and used the same model as described in Section \ref{onphase} but rescaled to the overall pulse phase.
\psrj\ is modeled as a power law with an exponential cutoff instead of a sub-exponential cutoff because of the lower statistics in each time bin.
Figure \ref{long_LC}, top panel shows the resulting long-term light curve of \psrj\/, which is well fitted by a constant line, yielding a reduced $\chi^{2}$ of {1.36}.
No significant flux variation is detected.
Three glitches have been detected from \psrj\/ (Table \ref{ephemeris}) and no flux variations are detected in the long-term light curve around the glitches.
We have checked for any changes in the spectrum of the pulsar around the glitches.
To accomplish this, the \emph{Fermi}-LAT data were split into four bins around the glitches.
Adopting the power law with exponential cutoff model, we performed a likelihood analysis for \psrj\ in these four time bins.
The flux and spectral parameters are shown in Figure \ref{long_LC} and are similar in the four epochs.
No change in the integral flux above 100 MeV is seen.

Adopting the best-fit spatial and spectral model derived from {the} above phase averaged analysis, we calculated the probabilities of photons
coming from \psrj\/ within a radius of 3$\degree$ using \textit{gtsrcprob} and produced a weighted pulsed light curve based on them.
The bottom panel of Figure \ref{profile} shows the folded, pulsed light curve above 100 MeV.
The remaining panels of the same figure show the light curve in narrower energy bands.
The light curve shows two distinct peaks, which is consistent with the profile reported by Abdo et al. (2012) and the 2PC.
To locate the two peaks, we fitted the light curve with a double Gaussian profile (Figure \ref{profile}).
The first (P1) and second (P2) peaks are at $\phi$=0.113$\pm$ {0.001} and $\phi$=0.293$\pm$ 0.001, respectively.
The separation between the means of the two peaks is 0.180$\pm$0.002.
The widths of P1 and P2 evolve with energy, leading to narrower peaks at higher energies (Figure \ref{gaussian_fit}, top and middle).
{A similar evolution} was also observed in Geminga (Abdo et al. 2010).
The relative strength of P1 and P2 decreases significantly from low to high energies (Figure \ref{gaussian_fit}, bottom panel), which again is consistent with what was first reported by Abdo et al. (2012).

\begin{center}
\begin{figure*}
\centering
\includegraphics[scale=0.6]{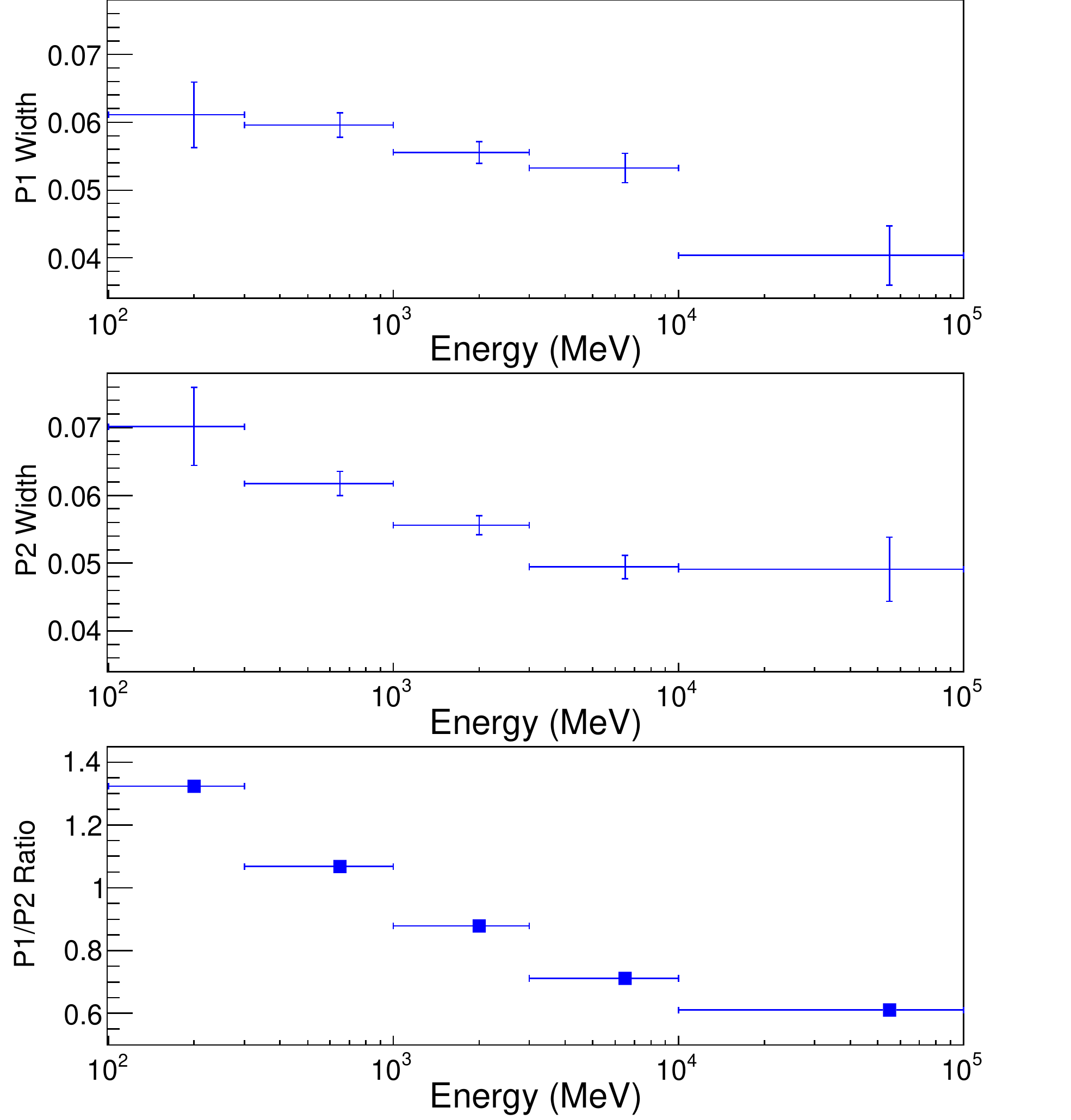}
\caption{From top to bottom, energy evolution of the width of P1, {the width of P2, and the P1/P2 ratio}.
The energy bins are the same as in Figure \ref{profile}.}
\label{gaussian_fit}
\end{figure*}
\end{center}

\section{DISCUSSION}
\label{discussion}

Using more than seven years of \emph{Fermi}-LAT data and a contemporaneous ephemeris, we carried out a detailed analysis of \psrj\ during its off-peak and its on-peak phase intervals.

During the off-peak phase, \psrj\ is significantly detected with a TS value of 262.
An exponential cutoff at 2.7 $\pm$1.2$\pm$ 1.3 GeV is {tentatively} detected in its spectrum, with a significance of 3.5 $\sigma$.
We explored the possible extension of \psrj\ during the off-peak phase, but a point-like source is favored (TS$_{ext}$=1.3).
The point-like nature of the emission together with the {potential} cutoff at GeV energies argue for a magnetospheric origin of the off-peak gamma-ray emission of \psrj\/.

Neither a point-like source nor extended gamma-ray emission was detected from \psrj\/ between 10 and 300 GeV during the off-peak phase.
{By removing the point source model of \psrj\/, assuming the same position and the 0.3-degree extension detected by VERITAS (Aliu et al. 2013), we calculated an upper limit for the possible emission coming from the PWN or the SNR CTA~1, of 6.5$\times$10$^{-12}$ erg cm$^{-2}$ s$^{-1}$ at 99\% confidence level, with Helene's method (Helene 1983) assuming a photon index of $2.0$ and considering the systematics {(10--300~GeV)}.
In the case of the highest energies, the TeV emission detected by VERITAS is most likely coming from the PWN.
The molecular mass in the vicinity of the complex is not enough to explain the TeV emission even under favorable assumptions for the cosmic-ray acceleration properties of the SNR (see the discussion by Martin et al. 2016).
The new upper limit we impose on the GeV emission from the PWN is not in conflict with detailed multi-frequency models (Aliu et al. 2013, Torres et al. 2014).
This PWN remains, however, a difficult case: it is unique in requiring a relatively high magnetization (as compared with other PWNe detected).
The latter and the SNR estimated age may indicate that the nebula (or at least part of it) is already contracting.
However, even considering that the PWN could already {have} passed reverberation, the needed magnetization is still high (Matin, Torres, \& Pedaletti 2016).

Off-peak emission of 26 young pulsars and 8 millisecond pulsars has been significantly detected (2PC).
Their off-peak luminosities range from $\sim$10$^{32}$ to $\sim$10$^{35}$ erg s$^{-1}$ and \psrj\ is near the geometric average (L$_{off\;peak}$= 3.5$\times$10$^{33}$ erg s$^{-1}$).
Considering a distance of 1.4 kpc and a spin-down power $\dot{E}$=$-I \Omega\dot{\Omega}$ ($I$ is the pulsar's moment of inertia$\sim$ 10$^{45}$ g cm$^{2}$, $\Omega$ and $\dot{\Omega}$ are the pulsar spin frequency and the first derivative of spin frequency)} of 4.5$\times$10$^{35}$ erg s$^{-1}$, the off-peak emission efficiency {(L$_{off\;peak}$/$\dot{E}$)} of \psrj\/ is $\sim$ 0.8\%, which is among the lowest of pulsars with magnetospheric off-peak emission (2PC, Figure 14).
The on-peak emission efficiency {(L$_{on\;peak}$/$\dot{E}$)} of \psrj\/ is $\sim$36.5\%.

For the on-peak phase, \psrj\ could be modeled by a power law with a sub-exponential cutoff, which is favored over an exponential cutoff with a significance above 11 $\sigma$ for the phase-averaged spectrum (Table \ref{psrj_fit}) and of 3 $\sigma$ for the phase-resolved spectra (Figure \ref{resolved}).
{This makes} \psrj\ the fourth pulsar having an established sub-exponential cutoff spectrum in at least some  phase range, besides Geminga, Vela, and Crab (see e.g., 2PC; Bochenek \& McCann 2015).
\psrj\/ showed a two-peak pulse profile.
The ratio of P1 and P2 evolves significantly with energy (Figure \ref{gaussian_fit}).
At low energies, {the strengths of P1 and P2 are} comparable, yet P2 is more prominent at higher energies (Figure \ref{profile}), similar to the tendency observed in the Crab, {Vela, and Geminga pulsars (Kanbach 1999}; Aleksi$\acute{c}$ et al. 2014).
This is consistent with the lower cut-off energy of P1 than that obtained for P2 (Figure \ref{resolved}, left panel).

Several hypotheses have been proposed to explain the deviation of the spectral cutoff from a pure exponential one.
In the outer-gap model of pulsar radiation, the high energy emission originates at high altitudes from the neutron star (see e.g., Cheng et al. 1986a, Cheng et al. 1986b, Romani 1996).
A spectral shape represented by a power law with an exponential cut-off is expected (see e.g., Prosekin et al. 2013, Vigan\`{o} et al. 2015a,b).
In these kinds of models, the accelerating electric field depends on the height in the gap (Hirotani 2006, Hirotani 2015).
Particles at distinct heights will be accelerated to different energies, leading to a range of cut-off energies.
The appearance of a sub-exponential cutoff could be taken as evidence that the emission of different pulsar phases {is produced by} different particle acceleration zones (or via different processes) with different radiation-reaction energies.
{As a result} of the wide emission beams in {the} outer-gap, {emission at a particular phase are a combination of different beams and different cutoff energies}.
{Therefore a} blend of different cutoff energies could plausibly lead to the sub-exponential spectra.
Leung et al. (2014) also proposed that the accelerating voltage in a given gap is unstable.
{Emission from even a single emitting zone is a convergence of various accelerate states,} which will lead to the sub-exponential cutoff in a pulsar spactra.
Such sub-exponential cutoffs can also be due to the contribution of a second component, arising from inverse Compton emission of electrons upscattering off soft photon fields (Hirotani 2015; Lyutikov 2013).
However, we note that the physical interpretation of the meaning of $b<1$ should be considered as provisional, since it may simply depend on our sensitivity.
{With increased statistics} we have seen that values of $b<1$ are needed to fit first the phase-averaged spectrum, then the phase-resolved ones.
It may well be that even in the smaller phase bins considered we are summing up contributions having different acceleration features and thus producing sub-exponential cutoffs as a result of this sum.
By reducing even further the phase bins, we would come to a situation in which cutoff power laws with $b=1$ and with $b<1$ would not produce significantly different fits.
Up to what extent the existence of $b<1$ is physical and not a problem of sensitivity (too large phase bins for the level of statistics attained) is still a subject of controversy.
For a phase-averaged analysis, we found no flux variability in the long-term light curve.
The integrated flux level and spectral parameters are consistent during all epochs preceding and following the glitches.

We have identified Fermi J0020+7328, a previously unknown, flaring gamma-ray source appearing (due to the relative strength of both sources) only during the off peak phases of \psrj\/.
The most probable counterpart for this source is \qso\/.

\acknowledgments

The \textit{Fermi} LAT Collaboration acknowledges generous ongoing support
from a number of agencies and institutes that have supported both the
development and the operation of the LAT as well as scientific data analysis.
These include the National Aeronautics and Space Administration and the
Department of Energy in the United States, the Commissariat \`a l'Energie Atomique
and the Centre National de la Recherche Scientifique / Institut National de Physique
Nucl\'eaire et de Physique des Particules in France, the Agenzia Spaziale Italiana
and the Istituto Nazionale di Fisica Nucleare in Italy, the Ministry of Education,
Culture, Sports, Science and Technology (MEXT), High Energy Accelerator Research
Organization (KEK) and Japan Aerospace Exploration Agency (JAXA) in Japan, and
the K.~A.~Wallenberg Foundation, the Swedish Research Council and the
Swedish National Space Board in Sweden.

Additional support for science analysis during the operations phase is gratefully acknowledged from the Istituto Nazionale di Astrofisica in Italy and the Centre National d'\'Etudes Spatiales in France.

We acknowledge the assistance from Dr. M. Kerr with the gamma-ray ephemeris for \psrj\/, Dr. M. Razzano and P. Saz Parkinson for discussions.
We acknowledge the support from the grants AYA2015-71042-P, SGR 2014-1073 and the National Natural Science Foundation of
China via NSFC-11473027, NSFC-11503078, NSFC-11133002, NSFC-11103020, XTP project XDA 04060604
and the Strategic Priority Research Program ``The Emergence of Cosmological Structures" of the Chinese Academy of Sciences, Grant No. XDB09000000.


\begin{thebibliography}{99}

\bibitem[Abdo A., et al. (2010)]{abdo2009} Abdo, A. A., Ackermann, M., Atwood, W. B., et al. 2008, Sci., 322, 1218
\bibitem[Abdo A., et al. (2010)]{abdo2009} Abdo, A. A., Ackermann, M., Ajello, M., et al. 2010, ApJ, 712, 1209
\bibitem[Abdo A., et al. (2010)]{abdo2009} Abdo, A. A., Ackermann, M., Ajello, M., et al. 2010, ApJ, 713, 154
\bibitem[Abdo A., et al. (2010)]{abdo2009} Abdo, A. A., Wood, K. S., DeCesar, M. E., et al. 2012, ApJ, 744, 146
\bibitem[Abdo A., et al. (2010)]{abdo2009} Abdo, A. A., Ajello, M., Allafort, A., et al. 2013, ApJS, 208, 17 (2PC)
\bibitem[Abdo A., et al. (2010)]{abdo2009} Acero, F., Ackermann, M., Ajello, M., et al. 2015, ApJS, 218, 23 (3FGL)
\bibitem[Abdo A., et al. (2010)]{abdo2009} Acero, F., Ackermann, M., Ajello, M., et al. 2016, ApJS, 223, 26
\bibitem[Abdo A., et al. (2010)]{abdo2009} Ackermann, M., Ajello, M., Allafort, A., et al. 2012, ApJS, 203, 4
\bibitem[Abdo A., et al. (2010)]{abdo2009} Ackermann, M., Ajello, M., Atwood, W. B., et al. 2015, ApJ, 810, 14
\bibitem[Abdo A., et al. (2010)]{abdo2009} Aleksi$\acute{c}$, J., Ansoldi, S., Antonelli, L. A., et al. 2014, A\&A, 565, 12
\bibitem[Abdo A., et al. (2010)]{abdo2009} Aliu, E., Archambault, S., Arlen, T., et al. 2013, ApJ, 764, 38
\bibitem[Abdo A., et al. (2010)]{abdo2009} Atwood, W. B., Abdo, A. A., Ackermann, M., et al. 2009, ApJ, 697, 1071
\bibitem[Abdo A., et al. (2010)]{abdo2009} Bertsch D. L., et al. 1992, Nature, 357, 306
\bibitem[Abdo A., et al. (2010)]{abdo2009} Bonnefoy, S.,  Brazier, K. T. S., Fichtel, C. E., et al. 2015, PoS, ICRC2015, 738
\bibitem[Abdo A., et al. (2009)]{abdo2009} Bochenek C., \& McCann A., 2015, proceeding of science of the 34th ICRC, arXiv:1507.03136
\bibitem[Abdo A., et al. (2009)]{abdo2009} Burrows D. N., Hill J. E., Nousek J. A., et al. 2005, Space Sci. Rev., 120, 165
\bibitem[Abdo A., et al. (2010)]{abdo2009} Caraveo, P. A., De Luca, A., Marelli, M., et al. 2010, ApJ, 725, L6

\bibitem[Abdo A., et al. (2010)]{abdo2009} Cheng, K. S., Ho, C., \& Ruderman, M. 1986a, ApJ, 300, 500
\bibitem[Abdo A., et al. (2010)]{abdo2009} Cheng, K. S., Ho, C., \& Ruderman, M. 1986b, ApJ, 300, 522


\bibitem[Abdo A., et al. (2009)]{abdo2009} de Jager, O. C., \& B$\ddot{u}$sching, I. 2010, A\&A, 517, L9
\bibitem[Abdo A., et al. (2009)]{abdo2009} de Jager, O. C., Raubenheimer, B. C., \& Swanepoel, J. W. H. 1989, A\&A, 221, 180
\bibitem[Abdo A., et al. (2010)]{abdo2009} Donato, D., Ghisellini, G., Tagliaferri, G.,\& Fossati, G., 2001, A\&A, 375, 739

\bibitem[Abdo A., et al. (2010)]{abdo2009} Halpern, J. P., Gotthelf, E. V., Camilo, F., Helfand, D. J., \& Ransom, S. M. 2004, ApJ, 612, 398
\bibitem[Abdo A., et al. (2010)]{abdo2009} Halpern, J. P., Camilo, C., \& Gotthelf, E. V. 2007, ApJ, 668, 1154
\bibitem[Abdo A., et al. (2010)]{abdo2009} Harris, D. E., \& Roberts, J. A. 1960, PASP, 72, 347
\bibitem[Abdo A., et al. (2010)]{abdo2009} Helene, O., 1983, NIMPR, 212, 319
\bibitem[Abdo A., et al. (2010)]{abdo2009} Hirotani, K. 2006, Modern Physics Letters A, 21, 1319
\bibitem[Abdo A., et al. (2010)]{abdo2009} Hirotani, K. 2015, ApJ, 798, L40
\bibitem[Abdo A., et al. (2010)]{abdo2009} Hobbs, G., Edwards, R., \& Manchester, R. 2006, Chin. J. Astron. Astrophys. Suppl., 6, 189
\bibitem[Abdo A., et al. (2010)]{abdo2009} Jackson, B., Scargle, J. D., Barnes, D., et al. 2005, ISPL, 12, 105
\bibitem[Abdo A., et al. (2009)]{abdo2009} Kanbach, G., 1999, ApL \& C, 38, 17
\bibitem[Abdo A., et al. (2009)]{abdo2009} Kerr, M. 2011, PhD thesis, Univ. Washington

\bibitem[Abdo A., et al. (2009)]{abdo2009} Kerr, M., Ray, P. S., Johnston, S., Shannon, R. M., Camilo, F., 2015, ApJ, 814, 128
\bibitem[Abdo A., et al. (2009)]{abdo2009} Lande, J., Ackermann, M., Allafort, A., et al., 2010, ApJ, 756, 5
\bibitem[Abdo A., et al. (2009)]{abdo2009} Lawrence, C. R., Pearson, T. J., Readhead, A. C. S., Unwin, S. C., 1986, AJ, 91, 494
\bibitem[Abdo A., et al. (2010)]{abdo2009} Leung,G. C. K, Takata, J., Ng, C. W., et al. 2015, ApJ, 797, 13
\bibitem[Abdo A., et al. (2010)]{abdo2009} Lin, L. C. C., Huang, R. H. H., Takata, J., et al. 2010, ApJ, 725, L1
\bibitem[Abdo A., et al. (2010)]{abdo2009} Lin, L. C. C., Hui, C. Y., Li, K. T., et al. 2014, ApJ, 793, 8
\bibitem[Abdo A., et al. (2010)]{abdo2009} Lyutikov, M., 2013, MNRAS, 431, 2580
\bibitem[Abdo A., et al. (2010)]{abdo2009} Mattox J. R., Bertsch, D. L., Chiang, J., et al., 1996, ApJ, 461, 396
\bibitem[Abdo A., et al. (2010)]{abdo2009} Martin J., Torres D. F., Pedaletti G. 2016, MNRAS, in press (astro-ph/1603.09328)
\bibitem[Abdo A., et al. (2010)]{abdo2009} McCann, A. 2015, ApJ, 804, 86
\bibitem[Abdo A., et al. (2010)]{abdo2009} Mignani, R. P., de Luca, A., Rea, N., et al. 2013 MNRAS, 430, 1354
\bibitem[Abdo A., et al. (2010)]{abdo2009} Pineault, S., Landecker, T. L., Madore, B., \& Gaumont-Guay, S. 1993, AJ, 105, 1060
\bibitem[Abdo A., et al. (2010)]{abdo2009} Pineault, S., Landecker, T. L., Swerdlyk, C. M., \& Reich, W. 1997, A\&A, 324, 1152

\bibitem[Abdo A., et al. (2010)]{abdo2009} Prosekin, A. Y., Kelner, S. R., \& Aharonian, F. A. 2013, arXiv: 1305.0783
\bibitem[Abdo A., et al. (2010)]{abdo2009} Ray, P. S., Kerr, M., Parent, D., et al. 2011, ApJS, 194, 17
\bibitem[Abdo A., et al. (2010)]{abdo2009} Romani, R. W., 1996, ApJ, 470, 469
\bibitem[Abdo A., et al. (2010)]{abdo2009} Sambruna, Rita M., 1997, ApJ, 487, 536
\bibitem[Abdo A., et al. (2010)]{abdo2009} Seward, F. D., Schmidt, B., \& Slane, P. 1995, ApJ, 453, 284
\bibitem[Abdo A., et al. (2010)]{abdo2009} Scargle, J. D., Norris, J. P., Jackson, B., \& Chiang, J. 2013, ApJ, 764, 167
\bibitem[Abdo A., et al. (2010)]{abdo2009} Slane, P., Seward, F. D., Bandiera, R., et al. 1997, ApJ, 485, 221
\bibitem[Abdo A., et al. (2010)]{abdo2009} Slane, P., Zimmerman, E. R., Hughes, J. P., et al. 2004, ApJ, 601, 1045
\bibitem[Abdo A., et al. (2010)]{abdo2009} Torres, D. F., Cillis, A., Mart\'in, J., \& de Ona Wilhelmi, E. 2014, JHEAp 1, 31

\bibitem[Abdo A., et al. (2010)]{abdo2009} Vigan\`{o}, D., \& Torres, D. F., 2015a, MNRAS, 449, 3755
\bibitem[Abdo A., et al. (2010)]{abdo2009} Vigan\`{o}, D., Torres, D. F., \& Mart\'in, J. 2015b, MNRAS, 453, 2599


\end{thebibliography}
\end{document}